\documentclass[10pt,aps,prx,twocolumn,superscriptaddress,floatfix,longbibliography]{revtex4-2}
\usepackage[T1]{fontenc}
\usepackage{graphicx}
\usepackage{amsmath,amssymb,bbm,amsthm}
\usepackage{enumerate}
\usepackage{braket}
\usepackage[colorlinks=true, allcolors=blue]{hyperref}
\usepackage[sort&compress]{cleveref}
\usepackage{xcolor}
\graphicspath{ {./Figures/} }
\providecommand{\abs}[1]{|#1|}
\providecommand{\norm}[1]{\|#1\|}
\newcommand{\opa}{\hat{a}}
\newcommand{\tr}{\operatorname{Tr}}
\newcommand{\vacuum}{0}
\newcommand{\lin}{\mathsf{L}}
\newcommand{\den}{\mathsf{D}}

\providecommand{\Cbb}{\mathbb{C}}

\providecommand{\Zbb}{\mathbb{Z}}
\providecommand{\T}{\mathsf{T}} 

\newcommand{\QQ}{\mathcal{Q}}

\newcommand{\CC}{\mathcal{C}}
\newcommand{\DD}{\mathcal{D}}
\newcommand{\EE}{\mathcal{E}}
\newcommand{\FF}{\mathcal{F}}
\providecommand{\Petz}[2]{\mathcal{R}^{\text{Petz}}_{#1, #2}}
\providecommand{\kb}[1]{\ket{#1}\!\bra{#1}}

\newcommand{\opD}{\hat{D}}

\newcommand{\opF}{\hat{F}}
\newcommand{\opG}{\hat{G}}

\newcommand{\opL}{\hat{L}}

\newcommand{\opS}{\hat{S}}

\newcommand{\opx}{\hat{x}}

\newcommand{\opR}{\hat{R}}

\newcommand{\opU}{\hat{U}}

\newcommand{\inp}[2]{\left \langle #1 \middle | #2 \right \rangle}

\newcommand{\uvec}{\mathbf{u}}
\newcommand{\vvec}{\mathbf{v}}
\newcommand{\fvec}{\mathbf{f}}
\newcommand{\cvec}{\mathbf{c}}
\newcommand{\alphavec}{\boldsymbol{\alpha}}

\newcommand{\Fchnl}{F_{\mathrm{chnl}}}
\newcommand{\cat}{\texttt{cat}}
\newcommand{\comb}{\texttt{comb}}

\begin{document}

\title{Passive Environment-Assisted Quantum Communication}

\author{Evelyn Voss}
\email{eavoss@uchicago.edu}
\author{Bikun Li}
\author{Zhaoyou Wang}

\author{Liang Jiang}
\affiliation{Pritzker School of Molecular Engineering, University of Chicago, Chicago, Illinois 60637}

\date{\today}

\begin{abstract}

As quantum information systems mature, efficient and coherent transfer of quantum information through noisy channels becomes increasingly important. We examine how passive environment-assisted quantum communication enhances direct quantum information transfer efficiency. A bosonic pure-loss channel, modeled as transmission through a beam splitter with a vacuum input state at the dark port, has zero quantum capacity when transmissivity is below 50$\%$. By using passive environment assistance—selecting an appropriate input state for the ancilla port—quantum communication through the channel can be enhanced. Although ideal Gottesman-Kitaev-Preskill (GKP) states enable perfect quantum information transmission at arbitrarily small transmissivity \cite{wang_passive_2025}, they remain challenging to realize with sufficient quality experimentally. We therefore explore more experimentally accessible non-Gaussian ancilla states, such as Fock, cat, and squeezed cat states, and numerically determine the optimal encoding and decoding strategies. We also construct analytical schemes that achieve high-fidelity transmission and good information rates.

\end{abstract}

\maketitle

\section{Introduction}

Quantum information processing ultimately hinges on the faithful transmission of quantum states through noisy channels. Determining how much information they can reliably convey and how to boost these limits remains a central problem in quantum information theory, as these channels govern the performance of communication protocols, error-correcting codes, and transduction interfaces.

The decoherence of quantum information in a channel arises primarily from two mechanisms: (i) energy loss, typically modeled as a bosonic pure-loss channel \cite{weedbrook_gaussian_2012} with transmissivity $\eta$ and (ii) interactions with the environment. One of the greatest challenges in quantum communication is overcoming the strict photon transmissivity $\eta$ requirement $\eta>0.5$. Below this threshold, no quantum information can be transferred \cite{wolf_quantum_2007}. The beam splitter formalism can be used to model the interaction between a bosonic system and its environment (importantly, this framework assumes a beam splitter with no intrinsic loss).
The quantum channels $\EE_1$ and $\EE_2$ describe the evolutions of the system and environment modes, $\opa_1$ and $\opa_2$, respectively. When the environment mode $\opa_2$ is initialized in the vacuum state, the induced system channel $\EE_1$ reduces to the bosonic pure-loss channel.

The performance of a quantum communication scheme can be quantified through multiple channel-based measures, including pre- and post-state fidelity \cite{han_microwave-optical_2021}, coherent information \cite{k_noh_quantum_2019}, and conversion bandwidth \cite{lauk_perspectives_2020}. Their relative importance depends on the intended application: overall efficiency $\eta$ is paramount for direct photon detection, whereas entanglement fidelity and coherent information rate are critical for distributed quantum computing and repeater-based communication \cite{lauk_perspectives_2020}, as they characterize both the quality of the transmitted quantum state and the rate at which quantum information can be reliably conveyed. Accordingly, we characterize our scheme using entanglement fidelity and coherent information rates.

Several known protocols have been developed to enhance quantum communication through noisy quantum channels. A standard approach, two-way quantum communication \cite{pirandola_fundamental_2017}, supplements the channel with bidirectional classical communication. Another line of work leverages environment control \cite{hayden_correcting_2005, winter_environment-assisted_2005, memarzadeh_recovering_2011, lami_bosonic_2020, mele_restoring_2022, mele_optical_2024, oskouei_capacities_2022}, motivated by the fact that any quantum channel can be realized as a unitary interaction $\hat U$ between an information-carrying system $S$ and an environment $E$ (Fig.~\ref{fig:BSfig}(b)). In active environment-assisted schemes, the environment is measured after the unitary and classical feedback is applied to the system \cite{hayden_correcting_2005, winter_environment-assisted_2005, memarzadeh_recovering_2011}. Both two-way communication and active assistance enable quantum information transmission through the bosonic pure-loss channel even at arbitrarily small transmissivity $\eta>0$ \cite{pirandola_fundamental_2017, hayden_correcting_2005}. For example, perfect transmission at any $\eta>0$ is achievable by preparing an infinitely squeezed vacuum in the environment, performing homodyne detection on the environment output, and applying inline squeezing to the system mode~\cite{zhang_quantum_2018}. In contrast, passive environment assistance, which is the focus of this work, requires only the preparation of a fixed environment state $\hat\sigma$ prior to the unitary interaction, avoiding the need for classical communication or feedback \cite{lami_bosonic_2020,mele_restoring_2022,mele_optical_2024}. 

In this work, we explore how passive environment-assisted quantum communication (PEAQC) can improve the direct quantum information transfer efficacy using experimentally accessible optical environment states. PEAQC exploits control over the preparation of the environment state and input state encoding to maximize the entanglement fidelity across the transduction scheme. Prior work has shown that preparing both the system input and the environment in suitable quantum error-correcting code states can significantly improve transmission rates \cite{wang_passive_2025}, a process framed in terms of encoding-decoding or encoding-recovery \cite{k_noh_quantum_2019, kosut_quantum_2009}.

It has been demonstrated that a Gottesman-Kitaev-Preskill or GKP-state encoding with GKP environment pairing is optimal for PEAQC and results in arbitrarily high entanglement fidelity across the scheme for almost any transmissivity \cite{wang_passive_2025}. Though there is much ongoing research on the efficient creation of GKP states, high-quality optical GKP states suitable for quantum communication are still not experimentally available \cite{konno_logical_2024, larsen_integrated_2025}. This motivates exploring alternative, currently realizable environment states and identifying their corresponding optimal encodings. In this work, we propose several PEAQC schemes employing non-GKP ancillary states that can be demonstrated with existing optical state-generation protocols.

High-fidelity preparation of low-photon-number Fock states has been demonstrated in both optical cavities and superconducting circuits \cite{hofheinz_generation_2008, cooper_experimental_2013, uria_deterministic_2020}. In addition to these states, we simulate cat environment states with amplitudes in the range $\alpha\in[1.5, 2.5]$ and squeezed-cat environment states with amplitudes in the range $\alpha\in[1.0, 1.5]$ and squeezing parameter $r = 0.7$. Recent experimental work has shown that cat states in this parameter regime are readily achievable with current techniques \cite{sun_generating_2023, he_fast_2023, luo_efficient_2024}. These experimentally feasible environment states form the basis of the PEAQC schemes explored in this work.

In practical scenarios, it is often necessary to control the environment of a quantum channel. Although passive environment assistance has been proposed for fiber communication \cite{lami_bosonic_2020, mele_restoring_2022, mele_optical_2024}, its implementation relies on leveraging nontrivial memory effects in optical fibers. We offer the application of quantum transduction as a more natural setting, in which the environmental modes of a transducer are directly accessible. As depicted in Fig.~\ref{fig:BSfig}(c), the transducer acts as a beam splitter that mixes the signal with an auxiliary environment input, allowing its performance to be enhanced through suitable environment-state preparation. Our analysis shows that passive environment assistance enables high-fidelity direct quantum transduction with the use of certain Fock, cat, and squeezed cat environment states even for transmissivities $\eta <0.5$.

\subsection{Summary of results}

In Section~\ref{sec:optimization}, we briefly review the computation methods for the encoding and decoding optimization program and present the corresponding results. Motivated by these optimization results, we next examine specific choices of encodings and environment states from an analytic standpoint. We begin with an analysis of the scheme with a cat-state environment in Section~\ref{sec:catstates}, followed by a characteristic-function treatment of a hexagonal GKP encoding paired with a Fock-state environment in Section~\ref{subsec:hexFock}. We then assess the performance of the scheme with Fock encoding and high Fock number state environment pairings in Section~\ref{subsec:highfock}.

\section{Optimizing quantum communication through a beam splitter}
\label{sec:optimization}

This section reviews the established biconvex optimization protocol developed in \cite{kosut_quantum_2009,fletcher_optimum_2007,albert_performance_2018} applied to our scheme with fixed Fock, cat, and squeezed cat environment states. We then explain the optimization results of our scheme with the Fock environment using characteristic functions.

 \subsection{Optimization scheme}

 \begin{figure}
    \includegraphics[width=1\linewidth]{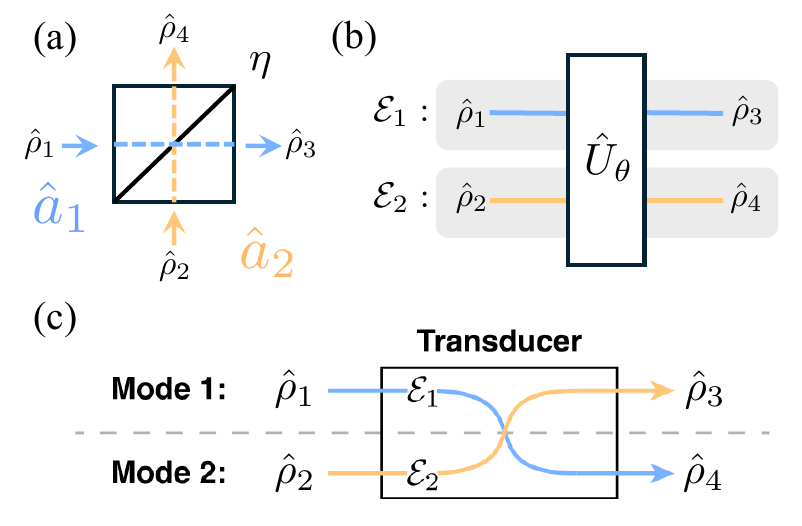}
    \caption{(a) Beam splitter with efficiency $\eta$. (b) Beam splitter unitary $\opU_\theta$ with quantum channels $\EE_1$ and $\EE_2$ that convert the input states $\hat{\rho}_1$ and $\hat{\rho}_2$ to output states $\hat{\rho}_3$ and $\hat{\rho}_4$, respectively. (c) Schematic of a two-mode transducer.}
    \label{fig:BSfig}
\end{figure}

The unitary operator for a beam splitter with efficiency $\eta \equiv \cos^2\theta$ ($\theta \in [0,\pi/2]$) is
\begin{align}
    \opU_\theta = \text{exp} \big[ \theta (\opa_1^{\dagger} \opa_2 - \opa_1 \opa_2^{\dagger} )\big]\;,
\end{align}
where $\opa_1$ and $\opa_2$ correspond to the system and environment modes, respectively. The system and environment input states are denoted by $\hat{\rho}_1$ and $\hat{\rho}_2$, respectively, while the corresponding output states are $\hat{\rho}_3$ and $\hat{\rho}_4$. Using the Schrödinger picture, the marginal output states are
\begin{align}
      \hat{\rho}_3 &= \EE_1 (\hat{\rho}_1) = \tr_2 \left[\opU_\theta (\hat{\rho}_1 \otimes \hat{\rho}_2) \opU_\theta^\dagger \right] \\
      \hat{\rho}_4 &= \EE_2 (\hat{\rho}_2) = \tr_1 \left[\opU_\theta (\hat{\rho}_1 \otimes \hat{\rho}_2) \opU_\theta^\dagger \right] .
 \end{align}
where the quantum channels $\EE_1 $ and $\EE_2$ describe the evolution of the modes as seen in Fig.~\ref{fig:BSfig}(b) and (c).

\begin{figure} [h]
    \centering
    \includegraphics[width=0.9\linewidth]{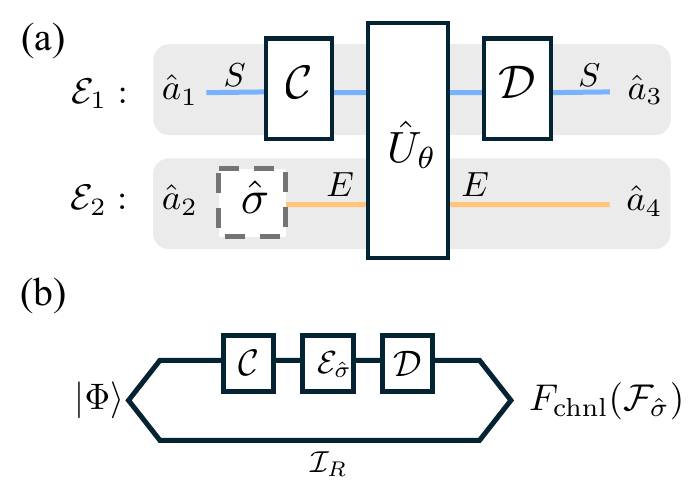}
    \caption{(a) Encoding/decoding scheme: the encoded input state $\CC(\hat{\rho})$ and environment state $\hat{\sigma}$ are acted upon by the unitary $\opU_\theta$, after which the system state is decoded by $\DD$. (b) Illustration of the entanglement fidelity calculation.}
    \label{fig:optscheme}
\end{figure}

To integrate these channels into the passive environment-assisted communication framework, we assign each mode pair (even or odd) to either the system or the environment subsystem. Let $\lin(\mathcal{H}_{S(E)})$ and $\den(\mathcal{H}_{S(E)})$ denote the respective spaces of linear operators and density operators acting on the system (environment) Hilbert space $\mathcal{H}_{S(E)}$. The resulting passive environment-assisted quantum channel $\EE_{\hat{\sigma}}: \den(\mathcal{H}_S) \to \den(\mathcal{H}_S)$ is defined as 
\begin{align}
    \EE_{\hat{\sigma}} (\hat{\rho}) = \tr_E \big[\opU_\theta (\hat{\rho} \otimes \hat{\sigma})\opU_\theta^\dagger\big]\;,
\end{align}
where $\hat{\sigma} \in \den(\mathcal{H}_E)$ is the environment state (Fig.~\ref{fig:optscheme}(a)). Note that $\EE_{\hat{\sigma}}$ reduces to the standard bosonic pure-loss channel $\EE_1$ when the environment is prepared in the vacuum state, $\hat{\sigma} = \ket{0}\bra{0}$.

Our goal is to faithfully transfer a qubit across this channel by encoding it into an optimal $n$-dimensional state within the channel Hilbert basis, sending it through the channel, and then decoding to retrieve the information as shown in Fig.~\ref{fig:optscheme}(a). Previous works have implemented this encoding-channel-decoding scheme with pure-loss and Gaussian thermal loss quantum channels \cite{wolf_quantum_2007}. For this work, we follow the same alternating semidefinite programming protocol as in \cite{kosut_quantum_2009, wang_passive_2025, fletcher_optimum_2007, albert_performance_2018}. Using the CVX package \cite{cvx_research_cvx_2012}, we find optimized encoding and decoding schemes for a PEAQC scheme initiated by random input encodings.

We assess the performance of a quantum channel using entanglement fidelity, which characterizes the degree to which the channel preserves entanglement with a reference system. For environment-assisted channels, the entanglement fidelity depends linearly on the choice of encoding, decoding, and environment state, making its maximization possible through an iterative convex optimization protocol \cite{k_noh_quantum_2019}.

The maximally entangled state on a $d$-dimensional input Hilbert space $\mathcal{H}_d$ and reference Hilbert space $\mathcal{H}_R$ is
\begin{align}
    \ket{\Phi} = \frac{1}{\sqrt{d}} \sum_{l = 0}^{d-1} \ket{l}\otimes\ket{l}_R\;,
\end{align}
where $\ket{l}$ ($\ket{l}_R$) are basis vectors for the $\mathcal{H}_d$ ($\mathcal{H}_R$) system.
Half of the maximally entangled state ($\ket{l}$) is sent through the encoding, channel, and decoding operations, while the other half ($\ket{l_R}$) is sent through an identity channel ($\mathcal{I}_R$) (Fig.~\ref{fig:optscheme}(b)). The entanglement fidelity can be expressed explicitly in terms of the full channel $\FF_{\hat{\sigma}} = \DD \circ \EE_{\hat{\sigma}} \circ \CC$ with encoding $\CC : \lin(\mathcal{H}_d) \to \lin(\mathcal{H}_S)$ and decoding $\DD: \lin(\mathcal{H}_S) \to \lin(\mathcal{H}_d)$ channels along with the environment state $\hat{\sigma}$ as

\begin{align}
    \Fchnl (\FF_{\hat{\sigma}}) := \bra{\Phi} \, \, \mathcal{I}_R \otimes \, (\DD \circ \EE_{\hat{\sigma}} \circ \CC)(\ket{\Phi}\bra{\Phi})\ket{\Phi}.
\end{align}

The quantum capacity $\QQ$ of the channel $\EE_{\hat{\sigma}}$ is lower-bounded by the single-letter quantum capacity of the channel, defined as 
\begin{align}
    \QQ^{(1)} (\EE_{\hat{\sigma}} ) := \max_{\hat{\rho}_1} I_c(\hat \rho_1,\EE_{\hat{\sigma}}),
    \label{eq:quantcap}
\end{align}
where the coherent information for our scheme given input state $\hat \rho$ and environment state $\hat{\sigma}$ is
\begin{align}
    I_c(\hat \rho,\EE_{\hat{\sigma}}) := S(\hat \rho_S) - S(\hat \rho_E),
    \label{eq:cohinfo}
\end{align}
given the system and environment output states, $\hat \rho_S := \tr_E[\FF_{\hat{\sigma}}(\hat \rho)]$ and $\hat \rho_E := \tr_S[\FF_{\hat{\sigma}}(\hat \rho)]$, respectively. We also define $S(\hat \rho)$ as the von Neumann entropy of a state $\hat \rho$. We thus use coherent information as an important metric of a quantum channel, establishing a lower bound on the channel's quantum capacity~\cite{wilde_quantum_2017}.

\subsubsection{Environment states}

A (2-leg) cat state is defined as $\ket{\cat_{2,\alpha}} = \frac{\ket{\alpha} + \ket{-\alpha}}{\sqrt{2}}$, where $\ket{\alpha} := \opD(\alpha)\ket{\vacuum}$, and $\opD(\alpha):=e^{\alpha \opa^\dagger - \alpha^* \opa}$ is the displacement operator.
We call the following definition of the logical $\ket{0_{L}}$ and $\ket{1_{L}}$ states the ``cat encoding'' $\CC_{\cat}:\lin(\Cbb^2)\to \lin(\mathcal{H}_S)$:
\begin{equation}
    \ket{0_{L}}:=\ket{\alpha},\quad
    \ket{1_{L}}:=\ket{-\alpha}\;.
\end{equation}
Note that this encoding is not strictly trace-preserving due to the small codewords' overlap. The output state will be normalized in our calculation.
A squeezed cat state can be written $\opS(r) \ket{\cat_{2,\alpha}}$ where the squeeze operator acting on mode-$i$ is defined as $\opS(r) = e^{\frac{r}{2}(\opa_i^2 - {\opa_i^{\dagger 2}})}$ ($r \in \mathbb{R}$). 

We ran our optimization scheme for $\eta \in [0,0.5]$ for Fock states $\ket{n}$ for $n=1,\ldots,4$, even cat states with amplitudes $\alpha = 1.5,2.0,2.5$, and squeezed cat states with amplitudes $\alpha=1.0,1.2,1.5$ and squeeze parameter $r = 1.4$.

\subsection{Optimization results}

\begin{figure*} [t]
    \centering
    \includegraphics[width=\linewidth]{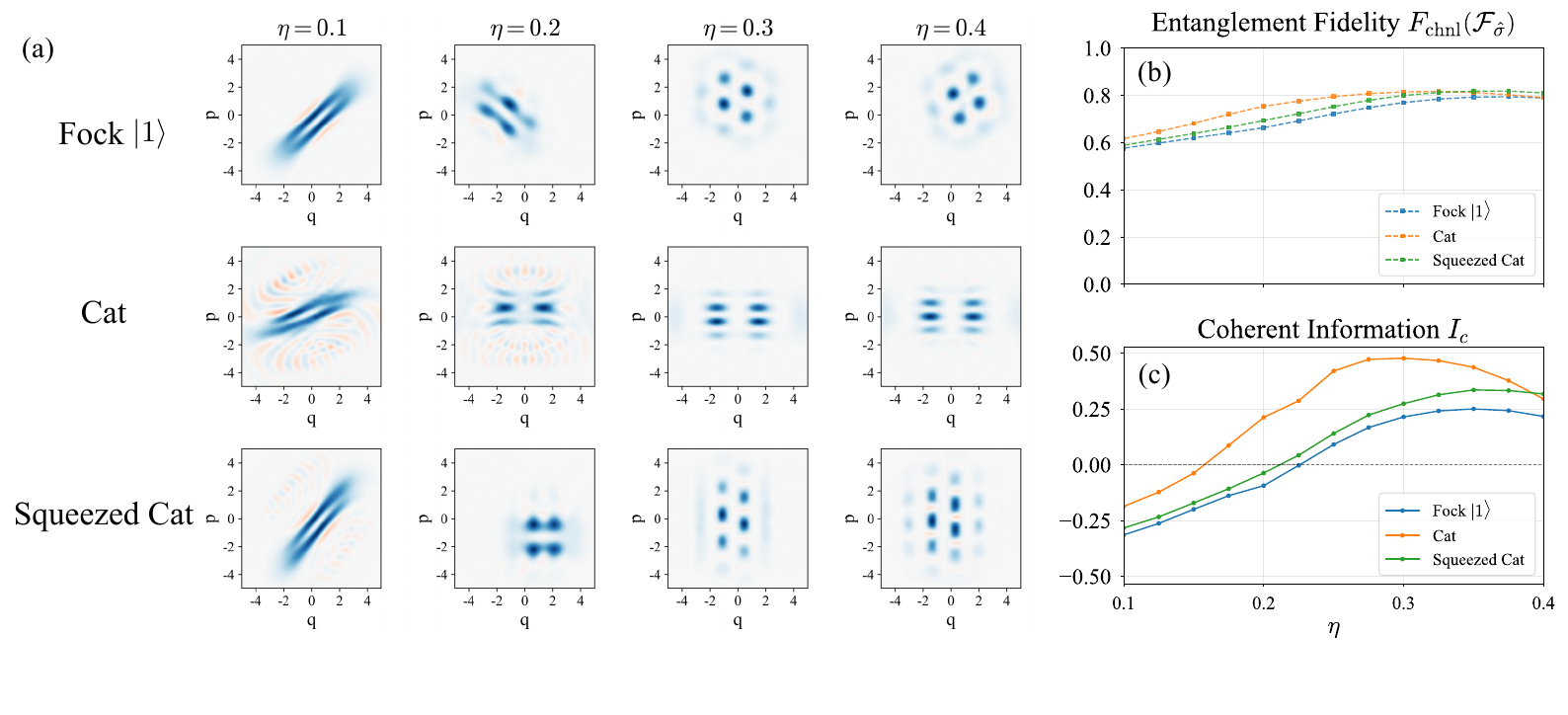}
    \caption{(a) Wigner functions of optimal encodings for a Fock $|1\rangle$ environment, cat state $\ket{\cat_{2,\alpha}}$ with $\alpha=2.0$, and squeezed cat state $\opS(r)\ket{\cat_{2,\alpha}}$ with amplitude $\alpha = 1.5$ and squeezing parameter $r = 1.4$. (b) Entanglement fidelity and coherent information rates for the channel (without decoding) with the Fock, cat, and squeezed cat environment states paired with their optimal encodings.}
    \label{fig:optresultsfig}
\end{figure*}

In Fig.~\ref{fig:optresultsfig} we present some of the results of our simulations with Wigner function representations of our optimal encodings, entanglement fidelities, and coherent information for each type of environment state. Using the protocol outlined above, the optimization converges after 150 rounds, where each round comprises 2 iterative steps (optimizing over encoding then decoding each round). We focus on nontrivial low transmissivities, $\eta\in[0.1, 0.4]$.

For the Fock-state environment $\lvert 1\rangle$, together with the encoding optimized for $\eta = 0.3$, we obtain an entanglement fidelity of $\Fchnl \approx 78\%$ and a coherent information of $I_c \approx 0.40$ (Fig.~\ref{fig:optresultsfig}(b)). The positivity of $I_c$ for the environment-assisted channel $\EE_{\hat{\sigma}}$ under this configuration certifies a nonzero quantum capacity. Consequently, passive environment assistance enables reliable quantum information transmission through $\EE_1$ at $\eta = 0.3$. By contrast, when the environment is prepared in the vacuum state, the quantum capacity of $\EE_1$ vanishes for all $\eta < 0.5$, so the channel alone cannot support quantum communication at this transmissivity~\cite{wilde_quantum_2017}.

The Wigner functions of the optimized encodings shown for Fock $|1\rangle$ environment, cat state $\ket{\cat_{2,\alpha}}$ (with $\alpha=2.0$), and squeezed cat state $\opS(r)\ket{\cat_{2,\alpha}}$ (with amplitude $\alpha = 1.5$ and squeezing parameter $r = 1.4$) all achieve positive coherent information rates for some value of $\eta < 0.5$ (Fig.~\ref{fig:optresultsfig}).

\section{PEAQC with cat and comb states}
\label{sec:catstates}
\begin{figure*}[t]
    \centering
    \includegraphics[width=\linewidth]{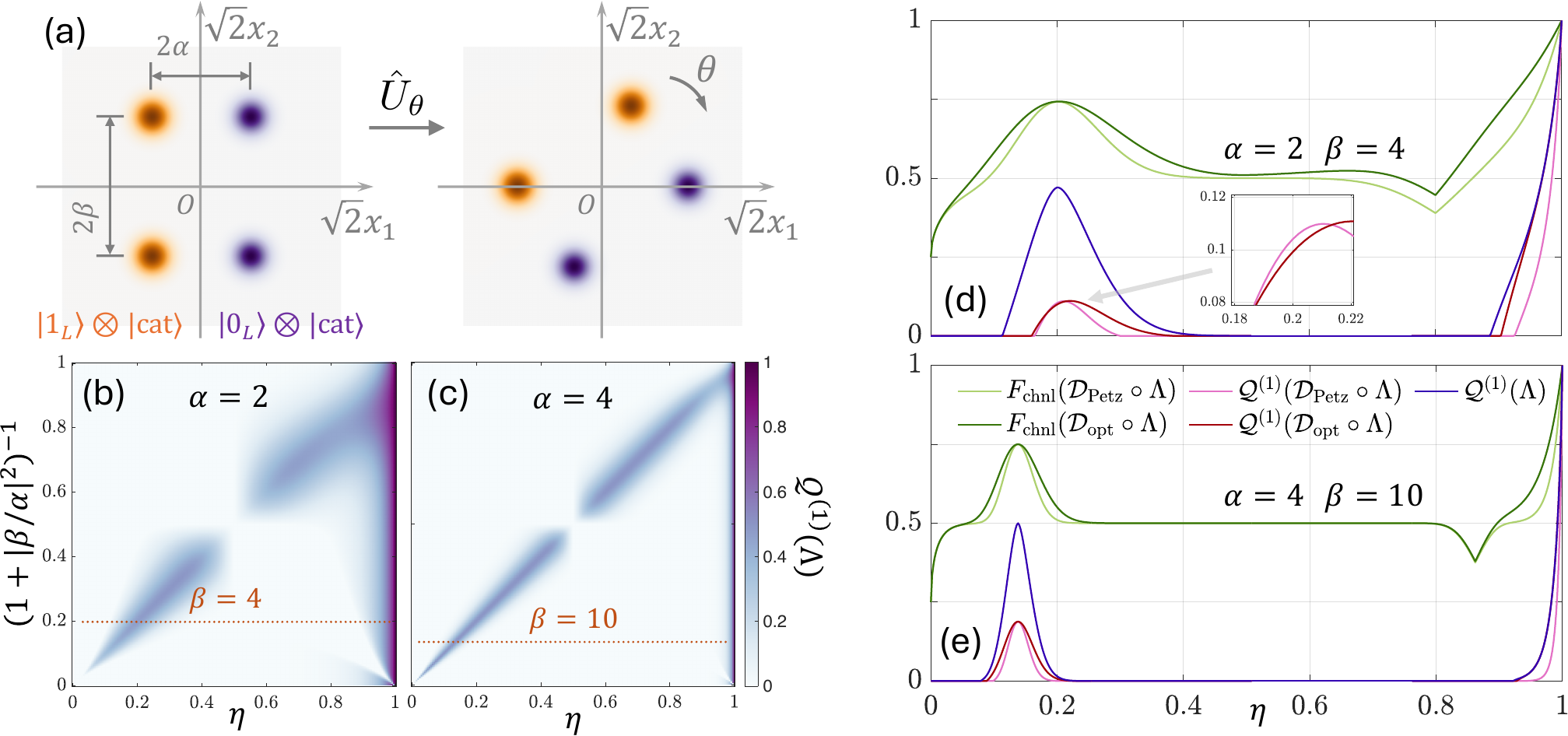}
    \caption{Panel (a) displays the 2D wavefunctions over real space $x_1$-$x_2$ for the global states $\ket{0_L}\otimes \ket{\mathtt{cat}_{2,\beta}}$ (purple) and $\ket{1_L}\otimes \ket{\mathtt{cat}_{2,\beta}}$ (yellow). The shading represents the amplitude of each wavefunction.
    The purified action of the loss channel is realized by the unitary $\opU_{\theta}$, which rotates the space clockwise by an angle $\theta = \arccos\sqrt{\eta} \in[0,\pi/2]$. Panels (b) and (c) show the single-letter quantum capacity $\QQ^{(1)}$ of $\Lambda\equiv\Lambda_{\cat,\eta}$ for different values of $\alpha$, $\beta$, and $\eta$. The vertical axis is rescaled as $(1+|\beta/\alpha|^2)^{-1}$. Panels (d) and (e) provide detailed views along the orange dotted line cuts in panels (b,c), showing the channel fidelities and single-letter quantum capacities for different decoders.}
    \label{fig:cat22}
\end{figure*}
This section explores the case where this non-Gaussian state is a cat state, as cat states are among the simplest and most experimentally well-explored non-Gaussian states~\cite{sun_generating_2023, he_fast_2023, luo_efficient_2024}. We then investigate generalizations to squeezed cat states and comb states, which eventually lead to the limit of GKP states.

We remark that the orthogonality of cat codewords is approximate for finite $\abs{\alpha}$, since $\inp{0_L}{1_L} = e^{-2|\alpha|^2}$. 
The quantum channel model in this section is constructed by setting mode~$1$ to a quantum state spanned by $\{\ket{0_{L}},\ket{1_{L}}\}:=\{\ket{\alpha},\ket{-\alpha}\}$, with the environment state $\hat{\rho}_2 = \cat = \ket{\cat_{2,\beta}}\bra{\cat_{2,\beta}}$.
For simplicity, we consider real parameters $1< \alpha<\beta$, so that the 2D wavefunction $\psi_\mu(x_1, x_2)$ for $\opU_\theta\ket{\mu_{L}}\otimes \ket{\beta}$ over the real space $x_1$-$x_2$ remains real (see Fig.~\ref{fig:cat22}(a)).
The resulting quantum channel is denoted as $\Lambda_{\eta,\cat} = \EE_{\cat}\circ\CC_{\cat}$.
We claim that when $\alpha\gg 1$, $\Lambda_{\eta,\cat}$ yields a nonzero quantum capacity $\QQ(\Lambda_{\eta,\cat})$  at $\theta = \arctan \frac{\alpha}{\beta}$, or equivalently, $\eta = \cos^2\theta= (1+\abs{\beta/\alpha}^2)^{-1}$. This implies that $\QQ>0$ for some arbitrarily small $\eta$ by increasing both $\alpha$ and $\beta/\alpha$.
This is again given by calculating the maximal coherent information (the single-letter quantum capacity as given in Eqs.~\eqref{eq:quantcap} and \eqref{eq:cohinfo}) sent through $\Lambda$ with respect to the input state: $\QQ^{(1)}(\Lambda) = \max_{\hat{\rho}_1}I_c(\hat{\rho}_1,\Lambda)$.

We show that $\QQ^{(1)}(\Lambda_{\eta,\cat})$ approaches $1/2$ at $\eta = (1+\abs{\beta/\alpha}^2)^{-1}$ for wide ranges of $\abs{\alpha}$ and $\abs{\beta}$ (see Fig.~\ref{fig:cat22}(b-e)). In the next subsection, we demonstrate that it is feasible to maintain this nonzero capacity in the limit $\eta\to 0$ and $\alpha,\beta\to \infty$.   
When a decoding map $\DD:\lin(\mathcal{H}_S)\to \lin(\Cbb^2)$ is applied to recover the logical space, we show that the single-letter quantum capacity remains finite (Fig.~\ref{fig:cat22}(d,e)). In this case, we can calculate the channel capacity for the composite linear map $\lin(\Cbb^2)\to \lin(\Cbb^2)$. In particular, we consider the optimal decoder $\DD_{\mathrm{opt}}$, which maximizes $\Fchnl$, and the Petz decoder $\DD_{\mathrm{Petz}}:=\Petz{\hat{\sigma}}{\Lambda_{\eta,\cat}}$ with a maximally mixed reference state $\hat{\sigma}$, which may be suboptimal in $\Fchnl$ but can be efficiently calculated. These \textit{single-mode} decoders decrease the coherent information but still yield non-vanishing values.
Note that the Choi matrix of $\Lambda_{\eta,\cat}$ has finite rank, so all numerical calculations can be performed without approximation within a relevant finite-dimensional subspace. Moreover, within this relevant subspace, we avoid the unbounded negative-powered operator issue mentioned in Ref.~\cite{Lami_2018}, which considers the continuous-variable Petz map. 
The detailed derivation for numerical calculations is provided in Appendix~\ref{app:catstates_calculation}.

\subsection{Optimization under infinite squeezing limit}
As shown in Fig.~\ref{fig:cat22}(d,e), at $\eta = (1+\abs{\beta/\alpha}^2)^{-1}$, the channel fidelity of the decoded channel is approximately $3/4$, and the single-letter quantum capacity (maximal coherent information) exhibits local maxima of approximately $\frac{1}{2}$ and $1-h_2(3/4)$, depending on whether the decoder is applied. Here, $h_2(x):= -x\log_2 x - (1-x)\log_2(1-x)$ is the binary entropy function. 
This subsection introduces an effective model with infinite squeezing that yields these exact values.

The action of a single-mode Gaussian unitary $\opG_i$ on a mode-$i$ annihilation operator is given by $\opG_i^\dagger \opa_i \opG_i = u_i \opa_i + v_i\opa_i^\dagger + \gamma_i$, with $|u_i|^2 - |v_i|^2 = 1$ ($u_i,v_i,\gamma_i\in\Cbb$).
If the two-mode unitary $\opU_\theta$ satisfies
\begin{equation}\label{eq:GGUGG_U}
    (\opG_1' \opG_2')^\dagger \opU_\theta (\opG_1  \opG_2) = \opU_\theta\;,
    \\
\end{equation}
for some $\opG_i$, $\opG_i'$ and $\theta \in (0,\pi/2)$, then the following equivalent conditions hold:
\begin{equation}\label{eq:GGUGG_U_conditions}
\begin{aligned}
u_1 &= u_2 = u_1' = u_2',\quad\\
v_1 &= v_2 = v_1' = v_2',\quad\\
\begin{pmatrix}
    \gamma_1'\\ \gamma_2'
\end{pmatrix}&=
\begin{pmatrix}
    \cos\theta & \sin\theta \\ 
    -\sin\theta & \cos\theta
\end{pmatrix}
\begin{pmatrix}
    \gamma_1\\ \gamma_2
\end{pmatrix}\;. 
\end{aligned}
\end{equation}
where $u_i',v_i',\gamma_i'$ are associated with $\opG_i'$.
A proof of the equivalence between Eq.~\eqref{eq:GGUGG_U} and Eq.~\eqref{eq:GGUGG_U_conditions} is provided in Appendix~\ref{app:proof_GGUGG_U}.
We remark that the condition in Eq.~\eqref{eq:GGUGG_U_conditions} is not necessary when $\theta = 0$ or $\pi/2$. In these fine-tuned trivial cases, the constraints on $u_i$ and $v_i$ can be relaxed.
Therefore, if $\opU_\theta$ is modified by $\opG_i$ with ``the same amount of squeezing'' ($u_1 = u_2$ and $v_1 = v_2$) and arbitrary displacements, the corresponding output of the modified quantum channel merely differs by a Gaussian unitary. This follows because the original $\opU_\theta$ can always be recovered by appending an appropriate $(\opG_1')^\dagger (\opG_2')^\dagger$ satisfying the constraints in Eq.~\eqref{eq:GGUGG_U_conditions} at the output, and $(\opG_2')^\dagger$ can be omitted since it is subsequently traced out.

This useful observation reveals an equivalence: the model remains unchanged when $\hat{\rho}_1$ and $\hat{\rho}_2$ separately undergo arbitrary and identical single-mode Gaussian operations.
For instance, the model in the previous section with cat states can be replaced by squeezed cat states with additional displacements. 
Let the squeezing operator be $\opS(r) = e^{\frac{r}{2}(\opa^2 - {\opa^{\dagger 2}})}$. We have
\begin{equation}
    \opS(r)\opD(\alpha e^{r}) = \opD(\alpha)\opS(r),\quad \alpha, r\in \mathbb{R}\;.
\end{equation}
Thus, when the unsqueezed cat states (encoding) in two modes involve large $\abs{\alpha}$ and $\abs{\beta}$, we can replace them with $r$-squeezed cat states with smaller displacements $\alpha_0 = \alpha e^{-r}$ and $\beta_0 = \beta e^{-r}$.

We now consider the infinite squeezing limit ($r\to \infty$ on both modes), while $\alpha_0$ and $\beta_0$ remain finite. 
We adopt this limit solely to facilitate the discussion and calculation of coherent information, channel fidelity, and decoders. 
The infinite squeezing limit is obviously not practical. In fact, it may not always be beneficial. One can compare Fig.~\ref{fig:cat22}(b) and (c): when a larger value of $\alpha$ is used (equivalent to larger squeezing), the value of $\QQ^{(1)}(\Lambda)$ is generally smaller. We expect that under the infinite squeezing limit, $\QQ^{(1)}$ is nonzero only on some specific values of $\eta$.
The following analysis under the infinite squeezing limit will be treated as an effective model which approximates the performance of finite squeezing.

Specifically, when each individual Gaussian function is infinitely squeezed, it converges to a Dirac delta function in the $x_i$-quadrature. This is particularly helpful because the small overlaps between wave packets can be neglected. In this case, for real $\alpha$, the state
\begin{equation}\label{eq:xxi}
    \ket{x}_{x_i}:=\opD(\alpha)\opS(r)\ket{\vacuum}_i\;\;(r\gg 1)
\end{equation}
\textit{approximates} the eigenstate of $\opx_i = \frac{1}{\sqrt{2}}\left(\opa_i + \opa_i^\dagger\right)$ with eigenvalue $x=\sqrt{2}\alpha$.
Here, the subscript ``$x_i$'' is used to indicate the $x$-quadrature (real space) parameterization associated with the $i$-th mode. 
Without loss of generality, we can treat the normalization condition as $\inp{x}{y}_{x_i}=\delta_{x,y}$, rather than $\delta(x-y)$, since we only consider countably many distinct $x$ coordinates.

\begin{figure}[t]
    \centering
    \includegraphics[width=0.72\linewidth]{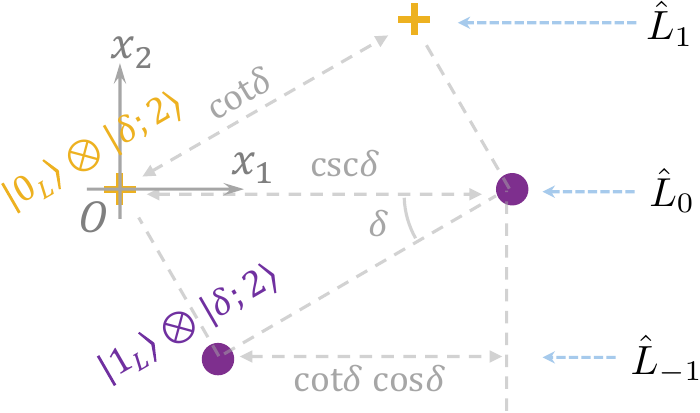}
    \caption{The markers (crosses, circles) represent the coordinates of the transformed infinitely squeezed Gaussian functions in Eq.~\eqref{eq:inf_sq_cat}. The rotation angle due to the beam splitter is $\theta = \pi/2 - \delta$.
    The environment state is denoted $\ket{\delta;2}$. Each Kraus operator $\opL_k$ corresponds to markers sharing the same $x_2$ coordinate.}
    \label{fig:comb_22}
\end{figure}

Hereafter, we adopt the notation in Eq.~\eqref{eq:xxi} with arbitrarily large $r$. Consider:
\begin{align}
\begin{split}\label{eq:inf_sq_cat}
    \ket{\mu_{L}}:=\ket{\mu}_{x_1},\quad
    \hat{\rho}_2:=\kb{\delta;2},\\ \ket{\delta;2}:=\frac{1}{\sqrt{2}}\left(\ket{0}_{x_2} + \ket{\cot\delta}_{x_2}\right)\;,
\end{split}
\end{align}
where $\mu\in \{0,1\}$ and $\delta\in(0,\pi/4]$.
Note that both $\ket{\mu_L}$ and $\ket{\delta;2}$ are shifted in real space compared to the previous cat state setting. Fig.~\ref{fig:comb_22} illustrates the layout of the 2D wavefunction.

In the large $r$ limit, the linear map $\Lambda$ is almost trace-preserving and can be considered a quantum channel with transmissivity $\eta$.
In the following analysis, we focus on the special case where $\eta$ and $\delta$ are related by $\eta = \sin^2\delta$. This quantum channel is denoted
\begin{equation}
\widetilde{\Lambda}_\delta:=\Lambda|_{\eta \equiv \sin^2\delta}\;.
\end{equation}
By calculating $\opU_\theta \ket{\mu_L}\otimes \ket{\delta;2}$, we find the following Kraus operators $\{\opL_k\}_{k=-1}^{1}$ for $\widetilde{\Lambda}_\delta$:
\begin{equation}\label{eq:Kraus_cat22_sq}
\begin{aligned}
    \opL_{-1} &= \sqrt{1/2}\ket{\sin\delta}_{x_1}\!\bra{1_L}\;,\\
    \opL_{0} &= \sqrt{1/2}\ket{0}_{x_1}\!\bra{0_L} + \sqrt{1/2}\ket{\csc\delta}_{x_1}\!\bra{1_L}\;,\\
    \opL_{1} &= \sqrt{1/2}\ket{\cot\delta\cos\delta}_{x_1}\!\bra{0_L}.\quad
\end{aligned}
\end{equation}
To obtain the above operators, we assume that all elements in $\{0,\csc\delta, \sin\delta, \cot\delta\cos\delta\}$ are distinct, which holds for \textit{almost} all values of $\delta$.
It follows that the coherent information is
\begin{align}
    I_c(\hat{\rho}_L, \widetilde{\Lambda}_\delta) &= S(\widetilde{\Lambda}_\delta(\hat{\rho}_L)) - S(\widetilde{\Lambda}_\delta^c(\hat{\rho}_L)) = \frac{S(\hat{\rho}_L)}{2}\;,
\end{align}
whose maximum $\max_{\hat{\rho}_L} I_c(\hat{\rho}_L, \widetilde{\Lambda}_\delta)=\QQ^{(1)}(\widetilde{\Lambda}_\delta)= \frac{1}{2}$ is achieved at $\hat{\rho}_L = \mathbbm{1}/2$. Since $\widetilde{\Lambda}_\delta$ is a degradable channel~\cite{wilde_quantum_2017}, we conclude that $\QQ(\widetilde{\Lambda}_\delta) = \QQ^{(1)}(\widetilde{\Lambda}_\delta) = 1/2$.

Based on $\opL_k$, further calculation of the ``QEC matrix $M$''~\cite{zheng_near-optimal_2024} yields
\begin{equation}\label{eq:pseudo_KLcond}
    M_{[\mu,k],[\nu,\ell]} := \bra{\mu_L}\opL_k^\dagger \opL_\ell \ket{\nu_L} = c^{(k)}_{\mu}\delta_{k\ell}\delta_{\mu\nu}\;,
\end{equation}
where $\cvec^{(-1)} = (0,1/2)$, $\cvec^{(0)} = (1/2,1/2)$, and $\cvec^{(1)} = (1/2,0)$.
This diagonal form does \textit{not} satisfy the exact Knill-Laflamme conditions but fulfills $[\sqrt{M}, \mathbbm{1}_L\otimes \tr_L\sqrt{M}] = 0$ (where the subscript ``$L$'' denotes the input logical subsystem). This vanishing commutator implies that the channel fidelity $\Fchnl$ for the optimally decoded channel $\DD_{\mathrm{opt}}$ can be achieved with the Petz map decoder $\DD_{\mathrm{Petz}} = 
\Petz{\hat{\sigma}}{\widetilde{\Lambda}_\delta}\big|_{\sigma\propto I}$~\cite{Iten2017TIT, li_optimality_2025}, where the reference state $\sigma$ is the maximally mixed state.  
By Ref.~\cite{zheng_near-optimal_2024}, this yields the optimal channel fidelity:
\begin{equation}
    \Fchnl = \frac{1}{d^2}\norm{\tr_L\sqrt{M}}^2_F = \frac{1}{d^2}\sum_k\norm{\cvec^{(k)}}_{\frac{1}{2}} = \frac{3}{4}\;,
\end{equation}
in which $\|\mathbf{v}\|_p := (\sum_i \abs{v_i}^p)^{1/p}$ is the vector $p$-norm of vector $\mathbf{v}$. The above value $\Fchnl$ can be seen in the green curves of Fig.~\ref{fig:cat22}(d,e), which provides the approximation result at finite $r$.
On the other hand, the Kraus operators for $\DD_{\mathrm{opt}} = \DD_{\mathrm{Petz}}$ are:
\begin{equation}\label{eq:RR_kraus_cat}
    \begin{aligned}
        \opR_{-1} &= \ket{1_L}\!\bra{\sin\delta}_{x_1}\;,\\
        \opR_{0} &= \ket{0_L}\!\bra{0}_{x_1} + \ket{1_L}\!\bra{\csc\delta}_{x_1}\;,\\
        \opR_{1} &= \ket{0_L}\!\bra{\cot\delta \cos\delta}_{x_1}.\\
    \end{aligned}
\end{equation}
It follows that the Kraus operators for the composite channel $\FF := \DD_{\mathrm{opt}}\circ \widetilde{\Lambda}_\delta$ are:
\begin{equation}\label{eq:FF_kraus_cat}
\begin{aligned}
    \opF_{-1} &= \sqrt{1/2}\kb{1_L},\\
    \opF_{0} &= \sqrt{1/2}(\kb{0_L} + \kb{1_L}),\\
    \opF_{1} &= \sqrt{1/2}\kb{0_L}.
\end{aligned}
\end{equation}
It follows that the coherent information for any input state $\hat{\rho}_L$ satisfies:
\begin{equation}
    \begin{aligned}
    &I_c(\hat{\rho}_L, \FF) \\ &= h_2\left(\frac{2+\sqrt{\norm{\mathbf{s}}_2^2+3s_z^2}}{4}\right)- h_2\left(\frac{2+\sqrt{1 + 3s_z^2}}{4}\right)\;, 
    \end{aligned}
\end{equation}
where $\hat{\rho}_L$ is represented by $\frac{1}{2}\left(\mathbbm{1}+s_x\hat{\sigma}_x + s_y\hat{\sigma}_y + s_z\hat{\sigma}_z\right)$ with $s_i\ge 0$ and Pauli matrix $\hat{\sigma}_i$. It is straightforward to verify that the maximum of $I_c$ is achieved when $\mathbf{s} = \mathbf{0}$, i.e., when the optimal $\hat{\rho}_L$ is maximally mixed, yielding $\QQ^{(1)}(\FF) = 1-h_2(3/4)\approx 0.1887$. Since $\FF$ (with Kraus representation Eq.~\eqref{eq:FF_kraus_cat}) is also a degradable channel \cite{wilde_quantum_2017}, we have quantum capacity $\QQ(\FF) = \QQ^{(1)}(\FF)$.

Finally, we emphasize that when the squeezing parameter $r$ is finite, the ``optimal decoder'' $\DD_{\mathrm{opt}}$ is optimal only in maximizing the channel fidelity, but is not guaranteed to optimize the decoded coherent information. As demonstrated in the inset of Fig.~\ref{fig:cat22}(d), the Petz map yields a higher maximal coherent information than the optimal decoder.

\subsection{Generalization: the comb states}
\begin{figure}[t]
    \centering
    \includegraphics[width=1\linewidth]{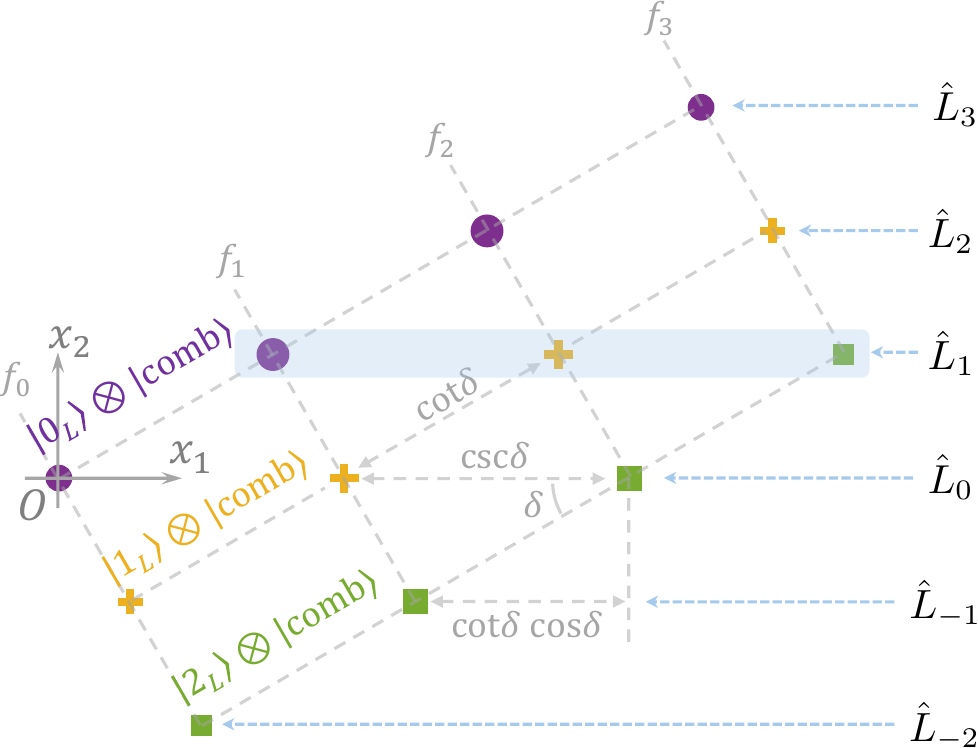}
    \caption{This figure demonstrates the generalization of Fig.~\ref{fig:comb_22}. The markers (circles, crosses, and squares) represent the coordinates of the transformed infinitely squeezed Gaussian functions in the comb state scheme (see Eq.~\eqref{eq:comb_scheme}). The rotation angle is $\theta = \pi/2 - \delta$.
    The environment state is denoted $\ket{\comb}\equiv \ket{\delta;m}$. In this example, $d=3$ and $m = 4$. The $x_1$ coordinate of each marker is given by $\xi(\mu,k;\delta)$ (Eq.~\eqref{eq:xi}). Each Kraus operator corresponds to markers sharing the same $x_2$ coordinate.}
    \label{fig:comb}
\end{figure}
The infinitely squeezed scheme in the previous section does not yield $I_c=1$ or $\Fchnl=1$. We now show that both fidelity and coherent information can be improved to their maximal values by generalizing the encoding and ancillary state to an infinitely squeezed ``comb state'' with appropriately chosen complex amplitudes $f_j$:
\begin{equation}\label{eq:comb_scheme}
  \begin{aligned}
    \ket{\mu_{L}}&:=\ket{\mu}_{x_1},\quad \mu\in \{0,1,\cdots,d-1\}\;,\\
    \ket{\delta;m}&:=
    \sum_{j=0}^{m-1}f_j\ket{j\cot\delta}_{x_2}\;,\quad m\ge d\;.
\end{aligned} 
\end{equation}
We call $\ket{\delta;m}$ the ``comb state'' $
\ket{\comb}$, whose complex amplitudes $f_j$ satisfy $\sum_{j} |f_j|^2 = 1$. 
For convenience, we set $f_j \equiv 0$ if $j<0$ or $j\ge m$.
We continue to use $\Lambda$ to denote the quantum channel (without decoder) for different values of $\eta$, under the comb state environment.

Again, following the notation from the last subsection, the quantum channel that transfers a qudit to a bosonic system with arbitrary $\eta$ is denoted $\Lambda$, while at $\eta = \sin^2\delta$, the quantum channel is denoted $\widetilde{\Lambda}_\delta := \Lambda|_{\eta \equiv \sin^2\delta}$.
Now $\widetilde{\Lambda}_\delta$ has a generalized set of Kraus operators (cf.\ Eq.~\eqref{eq:Kraus_cat22_sq}):
\begin{equation}\label{eq:Kraus_comb_sq}
\begin{aligned}
    \opL_k = \sum_{\mu = 0}^{d-1}f_{k+\mu}\ket{\xi(\mu,k;\delta)}_{x_1}\bra{\mu_L}\;,
\end{aligned}    
\end{equation}
with $k\in\{-d,-d+1,\cdots,0,\cdots,m-1\}$, 
where 
\begin{equation}\label{eq:xi}
    \xi(\mu,k;\delta) := \mu\csc\delta + k\cot\delta\cos\delta
\end{equation}
determines the coordinates associated with the output comb states. The derivation follows from the geometric relations shown in Fig.~\ref{fig:comb}.
Similarly, we consider the generic case where the values of $\xi(\mu,k;\delta)$ are distinct:
\begin{equation}\label{eq:distinct_xi}
    \inp{\xi(\mu,k;\delta)}{\xi(\nu,\ell;\delta)} = \delta_{\mu\nu}\delta_{k\ell}\;.
\end{equation}
This assumption yields a diagonal $M$, which satisfies the Petz map optimality condition in Eq.~\eqref{eq:pseudo_KLcond}. 
Thus, the optimal decoder $\DD_{\mathrm{opt}}$ is given by the Petz map with simple Kraus representation:
\begin{equation}\label{eq:RR_kraus_comb}
    \opR_k = \sum_{\mu = 0}^{d-1}\ket{\mu_L}\!\bra{\xi(\mu,k;\delta)}_{x_1}.
\end{equation}
Consequently, the Kraus operators for the composite channel $\FF=\DD_{\mathrm{opt}}\circ \widetilde{\Lambda}_\delta$ take a simple form:
\begin{equation}\label{eq:FF_kraus_comb}
\begin{aligned}
    \opF_k &= \sum_{\mu = 0}^{d-1}\abs{f_{k+\mu}}\kb{\mu_L}\;,
\end{aligned}
\end{equation}
where $k\in\{-d,-d+1,\ldots,m-1\}$ for both $\opR_k$ and $\opF_k$. We remark that Eq.~\eqref{eq:RR_kraus_comb} and \eqref{eq:FF_kraus_comb} are generalizations of Eq.~\eqref{eq:RR_kraus_cat} and \eqref{eq:FF_kraus_cat}. 

The channel fidelity for the above set of Kraus operators becomes a quadratic form in $|f_j|$:
\begin{equation}
    \Fchnl[d,\fvec] := \frac{1}{d^2}\sum_{k\in \Zbb}\left(\sum_{\mu=0}^{d-1}\abs{f_{k+\mu}}\right)^2.
\end{equation}
If $f_j$ is chosen to be $\sqrt{1/m}$ (and being zero for $j\notin\{0,1,\cdots,m-1\}$), we obtain $\lim_{m\to\infty} \Fchnl[d,\fvec] = 1$. It follows that the quantum capacity of $\widetilde{\Lambda}_\delta$ can approach $\log_2d$ for arbitrarily small transmissivity $\eta = \sin^2\delta$, provided $m$ is sufficiently large (which approximates the GKP state).
When $m$ is a finite integer, the maximum channel fidelity can be readily obtained via quadratic programming. Specifically, 
\begin{equation}
    \max_{\fvec}\Fchnl[d,\fvec]=\lambda_{\max}(\mathbf{T})/d^2,
\end{equation}
where $\lambda_{\max}(\mathbf{T})=d^2 - \mathcal{O}(m^{-2})$ is the maximal eigenvalue of an $m$-by-$m$ Toeplitz matrix:
\begin{equation}\label{eq:Toeplitz}
    \mathbf{T} = \begin{pmatrix}
        d&d-1&d-2&\cdots&0\\
        d-1&d&d-1&&\vdots\\
        d-2&d-1&d&\ddots&\\
        \vdots&&\ddots&\ddots&\\
        0&\cdots&&&d
    \end{pmatrix}\;.
\end{equation}
Here $\mathbf{T}$ has non-negative diagonals. For instance, the first row of $\mathbf{T}$ is $(3,2,1,0,0,\cdots,0)$, if $d = 3$.
In general, $\lambda_{\max}(\mathbf{T})$ and its corresponding eigenvector have no known closed-form expression in terms of $m$.
\begin{figure}[t]
    \centering
    \includegraphics[width=1\linewidth]{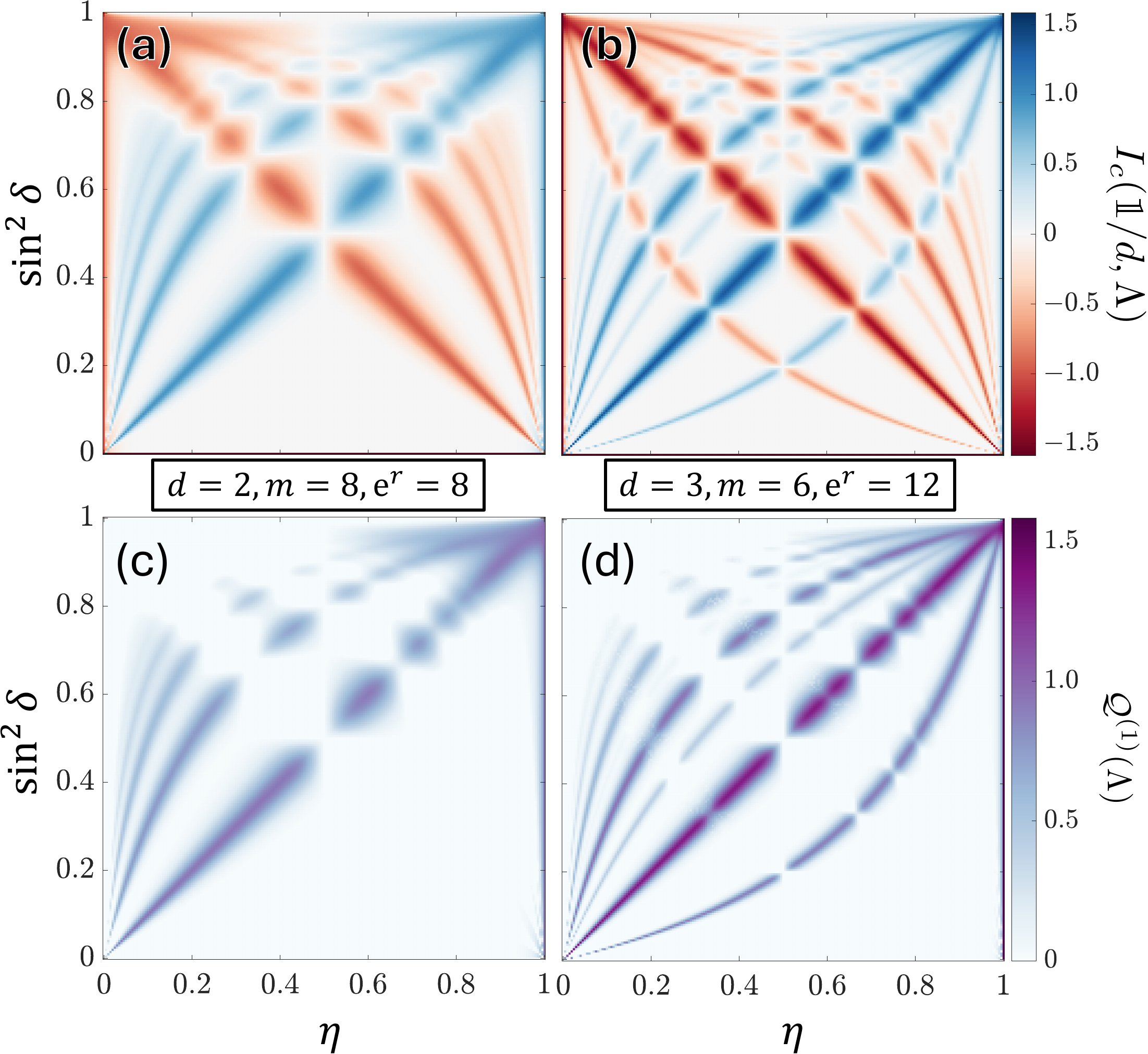}
    \caption{Heatmaps of $I_c(\mathbbm{1}/d,\Lambda)$ and $\QQ^{(1)}(\Lambda)$ for the general scheme in Eq.~\eqref{eq:comb_scheme} with finite squeezing parameter $r$. $\QQ^{(1)}(\Lambda)$ is always non-negative due to the optimization of the input state. Panels (a) and (c) correspond to $d=2$, $m=8$, and $e^r=8$. Panels (b) and (d) correspond to $d=3$, $m=6$, and $e^r=12$. The amplitude $f_i$ is given by Eq.~\eqref{eq:f_exact}.}
    \label{fig:heatmaps}
\end{figure}
However, when $d = 2$, the maximum channel fidelity has a compact form:
\begin{align}
\begin{split}
    \max_{\fvec}\Fchnl[2,\fvec] &= \frac{1}{4}\max_{\fvec}\sum_{j=0}^{m}\left(\abs{f_{j}} + \abs{f_{j-1}} \right)^2 \\ &= \cos^2\frac{\pi}{2(m+1)}\;,
\end{split}
\end{align}
which approaches $1$ for large $m$.
The above maximal channel fidelity is achieved when 
\begin{equation}\label{eq:f_exact}
    \begin{aligned}
        \abs{f_j} &= \sqrt{\frac{2}{m+1}} \sin \frac{(j+1)\pi}{m+1}\\j&\in\{0,1,\cdots,m-1\}\;.
    \end{aligned}
\end{equation}

Regarding the quantum capacities, we first consider the case of arbitrarily large $r$ for $\ket{x}_{x_i}$. As in the cat state scheme, from Eq.~\eqref{eq:FF_kraus_comb} we find that both $\DD_{\mathrm{opt}}\circ \widetilde{\Lambda}_\delta$ and $\widetilde{\Lambda}_\delta$ are degradable channels, so $\QQ = \QQ^{(1)}$.
Due to the simplicity of Eq.~\eqref{eq:FF_kraus_comb}, for $\FF := \DD_{\mathrm{opt}}\circ \widetilde{\Lambda}_\delta$, we find
\begin{equation}\label{eq:QQ1FF}
    \QQ^{(1)}(\FF) \ge I_c(\mathbbm{1}/d,\FF) = \log_2 d - S(\kappa) \;,
\end{equation}
where $S$ is the von Neumann entropy, and $\kappa\succeq 0$ is a $d$-by-$d$ trace-one matrix of the following component:
\begin{equation}
    \kappa_{ij} = \frac{1}{d}\sum_{k\in \Zbb} \abs{f_{k+i} f_{k+j}}\;.
\end{equation}
For $d=2$, $\kappa$ is a small matrix which allows us to optimize the lower bound in Eq.~\eqref{eq:QQ1FF}. One finds
\begin{equation}
    \begin{aligned}
        \max_{\fvec} I_c(\mathbbm{1}/2,\FF) &= 1 -h_2\left(\cos^2\frac{\pi}{2(m+1)}\right)
    \end{aligned}
\end{equation}
with the same optimal solution in Eq.~\eqref{eq:f_exact}.
For general $d\ge 2$, in the flat-amplitude case $f_j=\sqrt{1/m}$ for $j\in\{0,1,\cdots,m-1\}$, we have $S(\kappa)=\mathcal{O}(m^{-1}\log m)$, so the quantum capacity $\QQ(\FF)$ approaches $\log_2 d$ as $m$ increases.

In the comb state scheme, when the squeezing parameter $r$ for $\ket{x}_{x_i}$ is finite, Fig.~\ref{fig:heatmaps} displays heatmaps of the single-letter quantum capacity and the coherent information for maximally mixed state input. The amplitudes $f_j$ are determined from the eigenvector corresponding to $\lambda_{\max}(\mathbf{T})$ in Eq.~\eqref{eq:Toeplitz}. 
Fig.~\ref{fig:heatmaps}(a,c) shows the calculation for $d=2$, $m=8$, and Fig.~\ref{fig:heatmaps}(b,d) shows the calculation for $d=3$, $m=6$. As illustrated in panels (c) and (d), $\QQ^{(1)}$ approximates $\log_2 d$ for a wide range of $\eta = \sin^2\delta$, even when $e^r\approx 10$. Additionally, in panels (a) and (b), there is symmetry $I_c\to -I_c$ if $\eta\to 1-\eta$, regardless of $f_j$. The detailed derivation is provided in Appendix~\ref{app:I_coh_symm}.

Finally, we note that the comb state scheme (Eq.~\eqref{eq:comb_scheme}) can be generalized to the GKP scheme in Ref.~\cite{wang_passive_2025}. For instance, a GKP qubit encoding can be written as:
\begin{equation}
    \begin{aligned}
        \ket{0_L}_{\texttt{GKP}}&:=\cdots + \ket{0}_{x_1} + \ket{2}_{x_1} + \ket{4}_{x_1} + \cdots \\
        \ket{1_L}_{\texttt{GKP}}&:=\cdots + \ket{1}_{x_1} + \ket{3}_{x_1} + \ket{5}_{x_1} + \cdots \\
    \end{aligned}\;\;,
\end{equation}
which naturally follows from Eq.~\eqref{eq:comb_scheme} under appropriate displacement and squeezing. This scheme can be flexibly adjusted to many variants, which we leave to future work.

\section{PEAQC with Fock state environments}
\label{sec:Fock}
In this section, we explore schemes with Fock state environments.
We first provide an intuitive explanation of the numerical results where a hexagonal GKP code with Fock state environments (Fig.~\ref{fig:optresultsfig}) achieves good performance.
We then construct an analytical scheme with Fock states $\{\ket{0}, \ket{2}\}$ as the input encoding with different environment states. We construct a decoder and investigate bounds on the entanglement fidelity and quantum capacity of these channels.

\subsection{Hexagonal GKP encoding with Fock state environment}
\label{subsec:hexFock}

For a bosonic mode with creation and annihilation operators $\opa$ and $\opa^\dagger$, the quadrature operators are $\hat{x} = (\opa + \opa^\dagger)/\sqrt{2}$ and $\hat{p} = -i(\opa - \opa^\dagger)/\sqrt{2}$ where we have chosen the convention $\hbar
 = 1$.

A translation operator $\hat{T}$ acting on a quantum state in phase space is written as
\begin{align}
    \hat{T} (\uvec) \equiv \exp\{ i (u_p \hat{x} - u_x \hat{p}) \},
\end{align}
where $\uvec = (u_x,u_p) \in \mathbb{R}^2$ is the displacement vector by which the state is translated.

Note that $\hat{T}(\uvec) = \opD \left[(u_x + iu_p)/\sqrt{2}\right]$, where $\opD(\alpha)$ is the displacement operator as defined above in Sec.~\ref{sec:catstates}.
GKP states are defined as quantum states that remain invariant under specific phase-space displacements~\cite{gottesman_encoding_2001}. For a GKP state stabilized by the translations $\hat{T}(\uvec)$ and $\hat{T}(\mathbf{v})$, the corresponding stabilizer operators are $\opS_X = \hat{T}(\uvec)$ and $\opS_Z = \hat{T}(\mathbf{v})$. Ideal GKP states possess wavefunctions consisting of an infinite comb of delta functions, uniformly spaced along both the position and momentum axes. In phase space, these states therefore form a lattice.

The characteristic function $\chi_{\hat\rho}(\alphavec)$ of a quantum state $\hat{\rho}$ is defined as the expectation value of the Weyl-Heisenberg displacement operator (see Chapter 3 of \cite{walls_quantum_2025})
\begin{align}
    \chi_{\hat\rho} (\alphavec) = \tr\left[ \hat{\rho} \, \hat{T} (\alphavec)\right],
\end{align}
where $\hat{T}(\alphavec)$ is the phase-space translation operator with displacement vector $\alphavec = (\alpha_x, \alpha_p)$.

\begin{figure} [h]
    \centering
    \includegraphics[width=1\linewidth]{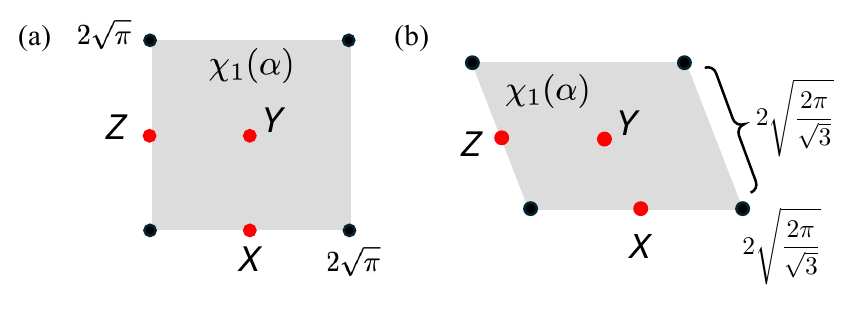}
    \caption{Unit cells of the characteristic function $\chi_1(\alpha)$ for (a) square-lattice and (b) hexagonal-lattice GKP states. The red dots indicate logical operator locations, while the black dots indicate stabilizer locations.}
    \label{fig:characfunc}
\end{figure}
\begin{figure} [h]
    \centering
    \includegraphics[width=1\linewidth]{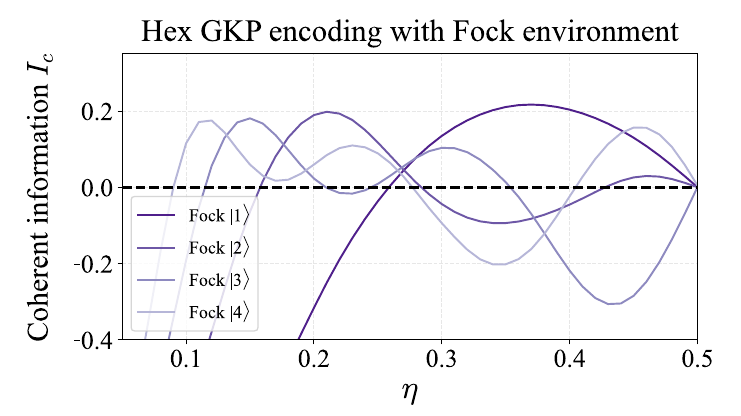}
    \caption{Coherent information of our scheme with the hexagonal GKP encoding with displacement vectors $u = \left(-\sqrt{\pi/4\sqrt{3}}, \sqrt{\pi\sqrt{3}/4}\right)$ and $v = \left(\sqrt{\pi/\sqrt{3}}, 0\right)$ paired with low Fock states $\ket{n}$ for $n = 1, \ldots, 4$}
    \label{fig:hexwithfockinfo}
\end{figure}

The characteristic function of an ideal GKP state forms a two-dimensional lattice in phase space, reflecting the periodic grid structure of the code in both quadratures. For a square GKP code, the stabilizers correspond to displacements by vectors
\begin{align}
    \uvec = (0, 2\sqrt{\pi}), \qquad \vvec = (2\sqrt{\pi}, 0),
\end{align}
and act on a logical code state $ \ket{\psi_L} = \alpha \ket{0_L} + \beta \ket{1_L} $ via
\begin{equation}
    \begin{aligned}
    \opS_q \ket{\psi_L} = \hat{T}(\uvec) \ket{\psi_L} = \ket{\psi_L}, \\
    \opS_p \ket{\psi_L} = \hat{T}(\vvec) \ket{\psi_L} = \ket{\psi_L}.
    \end{aligned}
\end{equation}

The logical Pauli operators correspond to half-stabilizer displacements:
\begin{align}
    \overline{Z} = \hat{T}(\uvec/2) = \hat{T}[(0, \sqrt{\pi})], \\
    \overline{X} = \hat{T}(\vvec/2) = \hat{T}[(\sqrt{\pi}, 0)].
\end{align}

Therefore, the characteristic function evaluated at the half-grid points in Fig.~\ref{fig:characfunc}(a) corresponds to the expectation values of logical Pauli measurements. Likewise, evaluation at the full stabilizer displacement vectors gives the stabilizer eigenvalues, which are equal to $+1$ for ideal GKP code states. Stabilizer measurements are marked by black dots, and logical measurements are marked by red dots.

To obtain a hexagonal GKP code, the lattice is tilted at an angle while preserving the area of the unit cell (see Fig.~\ref{fig:characfunc}(b)), giving rise to new displacement vectors
\begin{align}
    \uvec = \left(-\sqrt{2\pi/\sqrt{3}}, \sqrt{ 2\pi\sqrt{3}}\right), \quad \vvec = \left(2\sqrt{2\pi/\sqrt{3}},0\right),
\end{align}
and logical operators
\begin{align}
    \overline{Z} = \,& \hat{T}(\uvec/2) = \hat{T}\left(-\sqrt{\pi/2\sqrt{3}}\, , \, \sqrt{ \pi\sqrt{3}/2}\right), \\
    \overline{X} = \, & \hat{T}(\vvec/2) = \hat{T}\left(\sqrt{ 2\pi\sqrt{3}} \,, \, 0\right).
\end{align}

A beam splitter with efficiency $\eta$ acting on an input state $\hat{\rho} = \hat{\rho}_1 \otimes \hat{\rho}_2$ produces an output $\hat{\rho}'$,
whose marginal states $\hat{\rho}_3$ and $\hat{\rho}_4$ are described by the characteristic functions
\begin{equation}
\begin{aligned}
    \chi_3 (\alphavec) &= \chi_1(\sqrt{\eta} \, \alphavec) \chi_2 (\sqrt{1 - \eta} \, \alphavec)\;,
    \\
    \chi_4 (\alphavec) &= \chi_1(-\sqrt{1 - \eta} \, \alphavec) \chi_2(\sqrt{\eta} \, \alphavec),
\end{aligned}
\label{beamoncharac}
\end{equation}
where $\chi_k$ is the characteristic function of $\hat{\rho}_k$ for $k \in \{1,\cdots,4\}$.

For efficiencies $\eta =0.3$ and $0.4$, the optimal encoding for environment $\ket{1}$ appears to converge to a hexagonal GKP state (Fig.~\ref{fig:optresultsfig}(a)). We plot the coherent information for a sample hexagonal GKP state encoding paired with low Fock environments $\ket{n}$ for $n = 1, \ldots, 4$ in Fig.~\ref{fig:hexwithfockinfo}. For each low Fock environment state shown, this scheme achieves positive coherent information rates around $I_c\approx 0.2$ for some $\eta<0.5$.

\begin{figure} [h]
    \centering
    \includegraphics[width=0.9\linewidth]{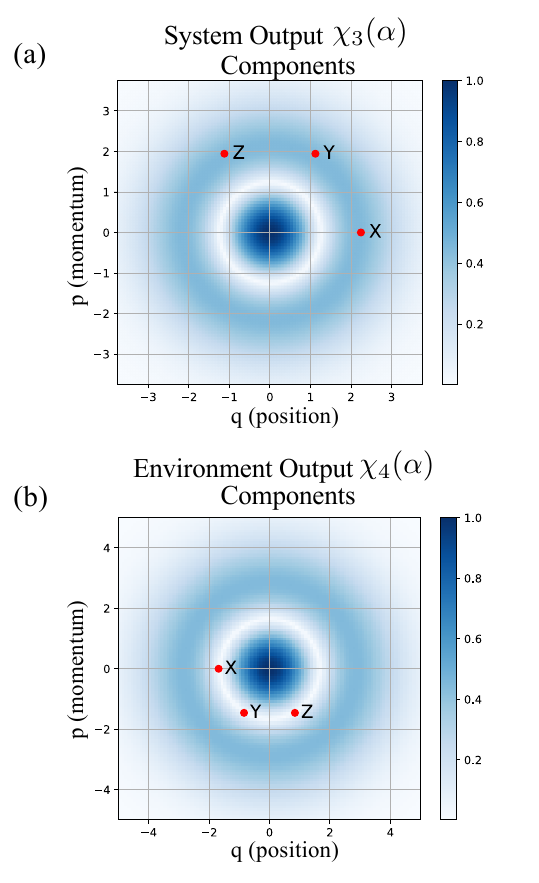}
    \caption{Characteristic function components for (a) the system output and (b) the environment output. The red dots correspond to logical operators encoded in the system. In (a), the system information is transmitted (rescaled $\chi_1(\alpha)$ according to Eq.~\eqref{beamoncharac}). In (b), the logical information is ``hidden'' by the blue environment amplitudes (rescaled $\chi_2(\alpha)$ according to Eq.~\eqref{beamoncharac}) due to the structure of the Fock state $\ket{1}$ at transmissivity $\eta = 0.34$.}
    \label{fig:charac_system_envf1}
    \vspace{-1\baselineskip}
\end{figure}
\vspace{30pt}

We illustrate in Fig.~\ref{fig:charac_system_envf1} the characteristic functions of Fock state environment $\ket{1}$. The characteristic function of a Fock state contains circular rings of zero amplitude, which can effectively ``hide'' the logical information of the input code state from the environment. In the system output [Fig.~\ref{fig:charac_system_envf1}(a)], the operators fall within a blue region in the system output, allowing the state $\hat{\rho}_3 \sim \chi_3(\alpha)$ to retain the logical information.
By contrast, in the environment output [Fig.~\ref{fig:charac_system_envf1}(b)], the logical operators (indicated by the red dots) of a hexagonal GKP code lie within a white region where $\chi_2(\sqrt{\eta}\alpha) = 0$, and thus are erased from $\hat{\rho}_4 \sim \chi_4(\alpha)$.

\begin{figure*} [t]
    \hspace*{-0.05\linewidth}
    \centering
    \includegraphics[width=1\linewidth]{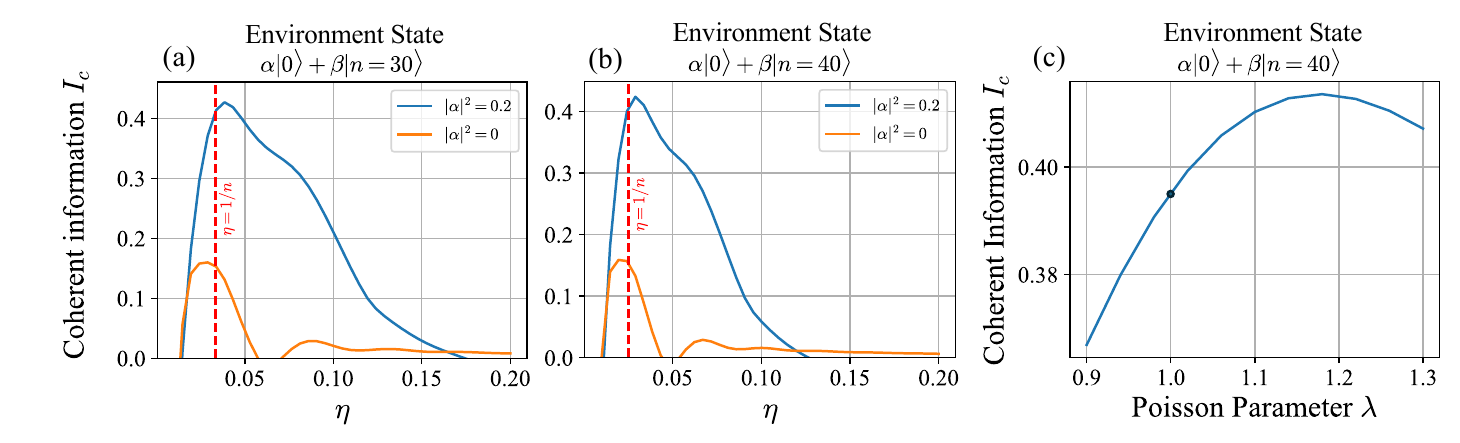}
    \caption{
     (a)/(b) Coherent information for a $\{\ket{0},\ket{2}\}$ encoding with an environment in the superposition state $\alpha\ket{0}+\beta\ket{n}$, plotted as a function of transmissivity $\eta$ for (a) $n=30$ and (b) $n=40$. The blue curves correspond to $|\alpha|^2=0.2$, while the orange curves represent a pure high-Fock environment state ($|\alpha|^2=0$). The superposition environment enhances the coherent information and yields a nonvanishing lower bound as $\eta\to0$ with $n\sim1/\eta$. (c) Coherent information lower bound for the $\{\ket{0},\ket{2}\}$ Fock encoding paired with a Fock $\alpha \ket{0}+\beta\ket{40}$ with optimized $|\alpha|^2$ for various choices of Poisson parameter $\lambda$.}
    \label{fig:FockFig}
\end{figure*}

\subsection{Fock \texorpdfstring{$\{\ket{0}, \ket{2}\}$}{{|0>, |2>}} encoding with high Fock environment states}
\label{subsec:highfock}
To transmit quantum information through a beam splitter at arbitrarily small $\eta>0$, we need to increase the number of photons in the input states.
Previous works~\cite{lami_bosonic_2020} have considered $\{\ket{0},\ket{1}\}$ encoding with Fock $\ket{n}$ environment states which achieve coherent information rates $I_c \approx0.1$.
Here we analyze modified schemes with $\{\ket{0},\ket{2}\}$ encoding with $\ket{\psi_E} = \alpha\ket{0}+\beta\ket{n}$ environment states that achieve much higher coherent information rates.

The performance of the $\{\ket{0},\ket{2}\}$ encoding with environment states $\ket{\psi_E} = \alpha\ket{0}+\beta\ket{n}$ for $n=30$ and $n=40$ is shown in Fig.~\ref{fig:FockFig}(a-c).
Having a nonzero vacuum component $\alpha$ in the environment state significantly increases the performance, and for $|\alpha|^2=0.2$ we get the highest coherent information rates of more than 0.4 near $\eta \sim 1/n$.

To better understand the scheme, we consider the limit of $\eta = \lambda/n$ where $\lambda$ is a constant as $n \to \infty$.
In this limit, we have the approximate output states
\begin{equation}
\begin{aligned}
\opU_\theta \ket{0}\ket{n}
&\approx \sum_{k=0}^{n} \sqrt{P_k}\,\ket{n-k}\ket{k}, \\
\opU_\theta \ket{2}\ket{n}
&\approx \sum_{k=0}^{n+2} \frac{\lambda}{\sqrt{2}} \sqrt{P_k}
\left(1-\frac{2k}{\lambda}+\frac{k(k-1)}{\lambda^2}\right)
\\&\quad\times
\ket{n+2-k}\ket{k},
\end{aligned}
\end{equation}
where the Poisson distribution
\begin{equation}
	P_k \equiv P_\lambda (k) = \frac{\lambda^k}{k!} e^{-\lambda}
\end{equation}
is the limit of binomial distribution.
Furthermore, we also have $\opU_\theta \ket{m_1} \ket{m_2} \approx (-1)^{m_1} \ket{m_2} \ket{m_1}$ for any $m_1,m_2 \ll n$.
In Fig.~\ref{fig:FockFig}(c), we optimize the coherent information over $|\alpha|^2$ for $\{\ket{0},\ket{2}\}$ encoding with $\alpha\ket{0}+\beta\ket{n}$ environment state, and plot the optimized results as a function of $\lambda$.
The optimal $\lambda$ is about $1.17$ with optimal coherent information about $0.415$.
Notice that the achievable coherent information rate only depends on $\lambda$ instead of $\eta$ and $n$. Therefore, we can achieve a capacity lower bound about 0.4 at arbitrarily small $\eta>0$ by choosing a large $n \approx \lambda / \eta$.

Below we analyze the special case of $\lambda = 1$ and derive a lower bound on the entanglement fidelity by explicit construction of a decoding channel.
For logical encoding $\{\ket{0}, \ket{2}\}$ and environment state $\ket{\psi_E} = \alpha \ket{\emptyset} + \beta \ket{n}$ ($\ket{\emptyset}$ can be any state whose photon numbers are much less than $n$ and $\ket{0}$ is one example), the output states are
\begin{equation}
	\begin{split}
		\opU_\theta \ket{0} \ket{\psi_E} \approx & \left( \alpha \ket{\emptyset} + \beta c \ket{n} \right) \ket{0} \\ &+ \beta c \ket{n-1} \ket{1} + \frac{\beta c}{\sqrt{2}} \ket{n-2} \ket{2} \\ & \qquad+ \frac{\beta c}{\sqrt{6}} \ket{n-3} \ket{3} \\
		\opU_\theta \ket{2} \ket{\psi_E} \approx & \frac{\beta c}{\sqrt{2}} \ket{n+2} \ket{0} - \frac{\beta c}{\sqrt{2}} \ket{n+1} \ket{1} \\ &+  \left( \alpha \ket{\emptyset} - \frac{\beta c}{2} \ket{n} \right) \ket{2} + \frac{\beta c}{2\sqrt{3}} \ket{n-1} \ket{3} ,
	\end{split}
\end{equation}
where $c = e^{-1/2}$ and we only keep $k=0,1,2,3$ terms since $P_{k>3} \approx 0$ for a Poisson distribution with $\lambda=1$.

To obtain a lower bound on the entanglement fidelity, we choose $\alpha,\beta$ such that $\alpha \ket{\emptyset} + \beta c \ket{n}$ and $\alpha \ket{\emptyset} - \frac{\beta c}{2} \ket{n}$ are orthogonal, which leads to $|\alpha|^2 = \frac{1}{2e+1}$. In this case, we can choose the decoding channel with Kraus operators
\begin{equation}
	\begin{split}
		\opR_0 &= \ket{0_L} \left(\sqrt{\frac{1}{3}} \bra{\emptyset} + \sqrt{\frac{2}{3}} \bra{n} \right) + \ket{1_L} \bra{n+2} \\
		\opR_1 &= \ket{0_L} \bra{n-1} - \ket{1_L} \bra{n+1} \\
		\opR_2 &= \ket{0_L} \bra{n-2} + \ket{1_L} \left(\sqrt{\frac{2}{3}} \bra{\emptyset} - \sqrt{\frac{1}{3}} \bra{n} \right) .
	\end{split}
\end{equation}
The corresponding entanglement fidelity (lower bound) is
\begin{equation}
	\Fchnl = \frac{\left( \sqrt{3}+1 \right)^2 + \left( \sqrt{2}+1 \right)^2 + \left( \sqrt{\frac{3}{2}} + 1 \right)^2}{4(2e+1)} \approx 0.71 .
\end{equation}

We also calculate the corresponding coherent information lower bound as $I_c \approx 0.39$ for $\lambda=1$, agreeing with the estimate from Fig.~\ref{fig:FockFig}(c).

\section{Discussion and outlook}
In this work, we have shown that passive environment-assisted quantum communication (PEAQC) can substantially enhance coherent information transmission through a bosonic pure-loss channel, even in regimes where the unassisted channel has zero quantum capacity. By pairing experimentally accessible non-Gaussian environment states with suitable encodings, we demonstrated positive coherent information for transmissivities well below $\eta=0.5$, without requiring idealized resources such as infinitely squeezed GKP states \cite{wang_passive_2025}.

A key theme of our results is that non-Gaussian structure in the environment can be exploited to suppress information leakage and protect logical degrees of freedom. For cat-state environments, we provided analytical constructions that yield nonzero single-letter quantum capacity at arbitrarily small transmissivity, supported by effective infinite-squeezing models. For Fock-state environments, characteristic-function analysis revealed how the structure of Fock states can ``hide'' logical information from the environment output for appropriate encodings, explaining the emergence of positive coherent information in our numerical optimizations.

Our analysis of Fock encodings paired with Fock environments further shows that relatively simple logical subspaces, such as the $\{\ket{0},\ket{2}\}$ encoding, already exhibit analytically tractable behavior. In particular, allowing the environment to occupy a superposition of vacuum and high–photon-number Fock states leads to a coherent information lower bound that remains finite as $\eta \to 0$, provided the environment energy scales as $n \sim 1/\eta$. This offers a concrete and experimentally motivated route to maintaining quantum communication capability in extreme-loss regimes.

Beyond their information-theoretic significance, these results are especially relevant for quantum transduction platforms. In many transducers, such as optomechanical, electro-optic, or microwave–optical interfaces, the auxiliary environmental modes are directly accessible and can be actively prepared, while the effective transmissivity is often well below 50\% \cite{mele_quantum_2022, mele_restoring_2022}. Our findings suggest that control over the environment state, combined with tailored bosonic encodings, can substantially enhance direct quantum state transfer and entanglement distribution across such devices without active feedback or measurement. Looking ahead, it would be valuable to extend this framework to multimode beam splitter models, which naturally arise in realistic quantum transducers involving several optical, mechanical, or microwave modes \cite{lauk_perspectives_2020}. Multimode interference and collective environment engineering may provide additional degrees of freedom for protecting logical information and enhancing coherent information beyond what is achievable in the two-mode setting.

Overall, this work highlights passive environment control as a powerful and experimentally realistic resource for quantum communication and transduction. By identifying low-complexity schemes that achieve nontrivial coherent information in high-loss regimes, we provide a foundation for near-term experimental demonstrations of environment-assisted quantum transduction and bosonic quantum networking.

\section{Acknowledgements}
We thank Ming Yuan, Debayan Bandyopadhyay, and Rohan Mehta for helpful discussions surrounding this project. We acknowledge support from the ARO (W911NF-23-1-0077), ARO MURI (W911NF-21-1-0325), AFOSR MURI (FA9550-21-1-0209, FA9550-23-1-0338), DARPA (HR0011-24-9-0359, HR0011-24-9-0361), NSF (ERC-1941583, OMA-2137642, OSI-2326767, CCF-2312755, OSI-2426975), and the Packard Foundation (2020-71479). 
This material is based upon work supported by the U.S. Department of Energy, Office of Science, National Quantum Information Science Research Centers and Advanced Scientific Computing Research (ASCR) program under contract number DE-AC02-06CH11357 as part of the InterQnet quantum networking project.

\bibliography{references}

\appendix
\onecolumngrid

\section{Complementary calculation for Sec.~\ref{sec:catstates}}
\subsection{The cat state scheme}\label{app:catstates_calculation}
Let us consider the cat state encoding $\CC_{\cat_{2,\alpha}}$ with a single Kraus operator:
\begin{equation}
    \hat{C} = \ket{\alpha}\bra{0_L} + \ket{-\alpha}\bra{1_L}\;.
\end{equation}
Note that this encoding is not trace-preserving, but we will still use the term ``quantum channel'' because the approximation error is exponentially small regarding $\abs{\alpha}$.
Before the unitary rotation $\opU_\theta = \exp[\theta(\opa_1^\dagger\opa_2 - \opa_1\opa_2^\dagger)]$, since the environment state is $\frac{\ket{\beta} + \ket{-\beta}}{\sqrt{2}}$, we have the initial state:
\begin{align}
    \ket{\psi_0}&:=\ket{0_L}\otimes \ket{\cat_{2,\beta}} = \frac{1}{\sqrt{2}}\ket{\alpha} \otimes \ket{\beta} + \frac{1}{\sqrt{2}} \ket{\alpha}\otimes \ket{-\beta}\;,\\
    \ket{\psi_1}&:=\ket{1_L}\otimes \ket{\cat_{2,\beta}} = \frac{1}{\sqrt{2}}\ket{-\alpha}\otimes \ket{\beta} + \frac{1}{\sqrt{2}}\ket{-\alpha}\otimes \ket{-\beta}\;.
\end{align}
To see the effect of $\opU_\theta$, we are interested in the transformation $\opU_\theta \ket{\alpha}\otimes \ket{\beta} = \ket{\alpha'}\otimes \ket{\beta'}$.
For convenience, let us define the non-orthogonal normalized states $\ket{a_{\mu j}}$ and $\ket{b_{\mu s}}$ ($\mu,j\in\{0,1\}$) by:
\begin{equation}
    \begin{pmatrix}
        a_{\mu j}(\theta) \\ b_{\mu j}(\theta)
    \end{pmatrix}
    :=
    R_\theta 
    \begin{pmatrix}
        (-1)^\mu\alpha \\ (-1)^j\beta
    \end{pmatrix}
    =
    \begin{pmatrix}
        (-1)^\mu\alpha \cos\theta + (-1)^j\beta\sin\theta\\
        -(-1)^\mu\alpha \sin\theta + (-1)^j\beta\cos\theta\\
    \end{pmatrix}\;,
\end{equation}
in which $R_\theta:=\begin{pmatrix}\cos\theta & \sin\theta \\ -\sin\theta & \cos\theta\end{pmatrix}$.
Thus, the action of $\opU_\theta$ can be represented as follows:
\begin{align}
    \opU_\theta\ket{\psi_0}&:= \frac{1}{\sqrt{2}}\ket{a_{00}} \otimes \ket{b_{00}} + \frac{1}{\sqrt{2}} \ket{a_{01}}\otimes \ket{b_{01}}\;,\\
    \opU_\theta\ket{\psi_1}&:= \frac{1}{\sqrt{2}}\ket{a_{10}}\otimes \ket{b_{10}} + \frac{1}{\sqrt{2}}\ket{a_{11}}\otimes \ket{b_{11}}\;.
\end{align}
It follows that we also obtain the following partially traced operators $(\mu,\nu \in \{0,1\})$:
\begin{align}
    J_{\mu\nu} &:= \tr_2\left(\opU_\theta\ket{\psi_\mu}\bra{\psi_{\nu}} \opU_\theta^\dagger\right) = \frac{1}{2}
    \begin{pmatrix}
        \ket{a_{\mu0}} & \ket{a_{\mu1}}
    \end{pmatrix}
    \begin{pmatrix}
         \inp{b_{\mu0}}{b_{\nu0}}&\inp{b_{\mu0}}{b_{\nu1}}\\
        \inp{b_{\mu1}}{b_{\nu0}}& \inp{b_{\mu1}}{b_{\nu1}}\\
    \end{pmatrix}
    \begin{pmatrix}
        \bra{a_{\nu0}} \\
        \bra{a_{\nu1}}
    \end{pmatrix}\;,\\
    J^c_{\mu\nu} &:= \tr_1\left(\opU_\theta\kb{\psi_\mu} \opU_\theta^\dagger\right) = \frac{1}{2}
    \begin{pmatrix}
        \ket{b_{\mu0}} & \ket{b_{\mu1}}
    \end{pmatrix}
    \begin{pmatrix}
         \inp{a_{\mu0}}{a_{\nu0}}&\inp{a_{\mu0}}{a_{\nu1}}\\
        \inp{a_{\mu1}}{a_{\nu0}}& \inp{a_{\mu1}}{a_{\nu1}}\\
    \end{pmatrix}
    \begin{pmatrix}
        \bra{b_{\nu0}} \\
        \bra{b_{\nu1}}
    \end{pmatrix}\;.
\end{align}
The (unnormalized) Choi state for the quantum channel $\Lambda_{\cat_{2,\beta}} = \EE_{\cat_{2,\beta}}\circ\CC_{\cat_{2,\alpha}}$ and its complementary channels can be expressed as follows
\begin{equation}\label{eq:choistates_cat22}
    J(\Lambda_{\cat_{2,\beta}}) = \begin{pmatrix}
        J_{00}&J_{01}\\
        J_{10}&J_{11}
    \end{pmatrix} \;,\quad
    J(\Lambda_{\cat_{2,\beta}}^c) = \begin{pmatrix}
        J^c_{00}&J^c_{01}\\
        J^c_{10}&J^c_{11}
    \end{pmatrix} \;.
\end{equation}
Note that the above Choi states have low-rank. Their output states are acting on Hilbert spaces with finite-dimensional supports $\operatorname{span}\{\ket{a_{\mu j}}\}$ and $\operatorname{span}\{\ket{b_{\mu\nu}}\}$. Therefore, if we are not interested in other degrees of freedom in the Fock space, we can define isometries:
\begin{align}
    \hat{V}_A&:=A (A^\dagger A)^{-\frac{1}{2}},\quad
    \hat{V}_B:=B (B^\dagger B)^{-\frac{1}{2}},\quad
\end{align}
with
\begin{align}
    A(\alpha,\beta,\theta) &:= \sum_{\mu,\lambda\in\{0,1\}}\ket{a_{\mu\lambda}}\bra{[\mu,\lambda]}\;,\\
    B(\alpha,\beta,\theta) &:= \sum_{\mu,\lambda\in\{0,1\}}\ket{b_{\mu\lambda}}\bra{[\mu,\lambda]}\;.
\end{align}
where the negative matrix power is calculated with the pseudo-inverse, and $\{\ket{[\mu,\lambda]}\}$ is a set of orthonormal states.
These notations allow us to represent the Choi states in Eq.~\eqref{eq:choistates_cat22} by the 8-by-8 matrices:
\begin{align}
    \widetilde{J}(\Lambda_{\cat_{2,\beta}}) &:= \big(\mathbbm{1}\otimes \hat{V}_A^\dagger\big) J(\Lambda_{\cat_{2,\beta}})\big(\mathbbm{1}\otimes \hat{V}_A\big)\;, \label{eq:finite_choi_cat22}\\
    \widetilde{J}(\Lambda_{\cat_{2,\beta}}^c) &:= \big(\mathbbm{1}\otimes \hat{V}_B^\dagger\big) J(\Lambda_{\cat_{2,\beta}}^c)\big(\mathbbm{1}\otimes \hat{V}_B\big)\;.\label{eq:finite_choi_comp_cat22}
\end{align}

Now we have all the utilities to find the coherent information $I_c(\hat{\rho}_L, \Lambda_{\cat_{2,\beta}})$ with finite matrices. 
Recall that the coherent information is $S(\Lambda_{\cat_{2,\beta}}(\hat{\rho})) - S(\Lambda_{\cat_{2,\beta}}^c(\hat{\rho}))$, where the output states are
\begin{align}
    \Lambda_{\cat_{2,\beta}}(\hat{\rho}_L) &= \tr_1 \left[(\hat{\rho}_L^\T\otimes \mathbbm{1}) J(\Lambda_{\cat_{2,\beta}})\right] = \sum_{\mu,\nu\in\{0,1\}} (\hat{\rho}_L)_{\mu\nu} J_{\mu\nu}\;,\\
    \Lambda_{\cat_{2,\beta}}^c(\hat{\rho}_L) &= \tr_1 \left[(\hat{\rho}_L^\T\otimes \mathbbm{1}) J(\Lambda_{\cat_{2,\beta}}^c)\right] = \sum_{\mu,\nu\in\{0,1\}} (\hat{\rho}_L)_{\mu\nu} J^c_{\mu\nu}\;.
\end{align}
The original infinite-dimensional Choi matrices have low rank. Thus, they can be replaced by a finite-dimensional representation in Eq.~\eqref{eq:finite_choi_cat22} and \eqref{eq:finite_choi_comp_cat22}. It follows that the expression of coherent information can be written as:
\begin{equation}\label{eq:Icoh_rho_Lambda}
    I_c(\hat{\rho}_L, \Lambda_{\cat_{2,\beta}}) = -\tr(\hat{\omega}_A\log_2 \hat{\omega}_A)+\tr(\hat{\omega}_B\log_2 \hat{\omega}_B)\;,
\end{equation}
where
\begin{align}
    \hat{\omega}_A(\hat{\rho}_L,\theta,\alpha,\beta) &:=   \frac{1}{\mathcal{N}_A}\sum_{\mu,\nu\in\{0,1\}} (\hat{\rho}_L)_{\mu\nu} \hat{V}_A^\dagger J_{\mu\nu} \hat{V}_A \;,\\
    \hat{\omega}_B(\hat{\rho}_L,\theta,\alpha,\beta) &:=  \frac{1}{\mathcal{N}_B}\sum_{\mu,\nu\in\{0,1\}} (\hat{\rho}_L)_{\mu\nu} \hat{V}_B^\dagger J^c_{\mu\nu} \hat{V}_B 
\end{align}
are 4-by-4 matrices. 
We highlight that $\CC_{\cat_{2,\alpha}}$ is generally trace-decreasing, so the above output states have traces slightly less than $1$. A normalization factor $\mathcal{N}_A(B)$ on $\hat{\omega}_A(B)$ is needed before sending them into the von Neumann entropy.
The simple form of Eq.~\eqref{eq:Icoh_rho_Lambda} allows efficient numerical calculations.

The above derivation provides a finite-dimensional representation of the complementary channel Choi state $J(\Lambda_{\cat_{2,\beta}}^c)$. Particularly, from Eq.~\eqref{eq:finite_choi_comp_cat22}, the ``QEC matrix'' $M$ of $\Lambda_{\cat_{2,\beta}}$ can be written as
\begin{equation}
    M = \widetilde{J}(\Lambda_{\cat_{2,\beta}}^c)^\T\;,
\end{equation}
from which one can obtain the channel fidelity under the Petz map decoder (with maximally mixed reference state) $\Fchnl(\DD_{\text{Petz}}\circ \Lambda_{\cat_{2,\beta}}) = \frac{1}{4}\|\tr_L\sqrt{M}\|_F^2$. To obtain the optimal decoder $\DD_{\text{opt}}$ that optimizes the channel fidelity, one can perform the SDP optimization with the finite matrix $\widetilde{J}(\Lambda_{\cat_{2,\beta}})$, as we only need to decode the quantum state within $\operatorname{span}\{\ket{a_{\mu j}}\}$.

\subsection{The comb state scheme}\label{app:combstates_calculation}
We now generalize the cat encoding to the comb encoding $\CC_{\comb}$, defined by a single Kraus operator:
\begin{equation}
    \hat{C} = \sum_{\mu = 0}^{d-1}\ket{\mu \alpha}\bra{\mu_L}\;.
\end{equation}
Note that we do not consider infinite squeezing in this section.
Let the environment state be $\sum_{j=0}^{m-1} f_j\ket{j\beta}$. The generalized global states are then
\begin{align}
    \ket{\psi_\mu} &:= \sum_{j = 0}^{m-1} f_j \ket{\mu\alpha}\ket{j\beta}\;,
    \quad
    \opU_\theta \ket{\psi_\mu} := \sum_{j = 0}^{m-1} f_j \ket{a_{\mu j}}\ket{b_{\mu j}}\;,
\end{align}
with $\mu \in \{0,1,\cdots, d-1\}$, and
\begin{equation}
    \begin{pmatrix}
        a_{\mu j} \\ b_{\mu j}
    \end{pmatrix}
    :=
    R_\theta
    \begin{pmatrix}
        \mu \alpha \\ j\beta
    \end{pmatrix}
    =
    \begin{pmatrix}
        \mu \alpha \cos\theta + j\beta \sin \theta\\
        -\mu \alpha \sin\theta + j\beta \cos \theta
    \end{pmatrix}
    \;.
\end{equation}
The Choi matrix subblocks and full Choi matrices for the quantum channel $\Lambda_{\comb}:=\EE_{\comb}\circ\CC_{\comb}$ are:
\begin{align}
    J_{\mu\nu}&:=\tr_2\left(\opU_\theta\ket{\psi_\mu}\bra{\psi_\nu}\opU_\theta^\dagger\right) = \sum_{j,k = 0}^{m-1} f_j f_k^* \ket{a_{\mu j}}\inp{b_{\mu j}}{b_{\nu k}}\bra{a_{\nu k}} \;,\label{eq:J_munu_comb}\\
    J_{\mu\nu}^c&:=\tr_1\left(\opU_\theta\ket{\psi_\mu}\bra{\psi_\nu}\opU_\theta^\dagger\right) = \sum_{j,k = 0}^{m-1} f_j f_k^* \ket{b_{\mu j}}\inp{a_{\mu j}}{a_{\nu k}}\bra{b_{\nu k}} \;,\label{eq:J_munu_comb_c}\\
    J(\Lambda_{\comb}) &= \begin{pmatrix}
        J_{00}&\cdots& J_{0,d-1}\\
        \vdots&\ddots&\vdots\\
        J_{d-1,0}&\cdots&J_{d-1, d-1}
    \end{pmatrix},\quad
    J(\Lambda_{\comb}^c) = \begin{pmatrix}
        J_{00}^c&\cdots& J_{0,d-1}^c\\
        \vdots&\ddots&\vdots\\
        J_{d-1,0}^c&\cdots&J_{d-1, d-1}^c
    \end{pmatrix}
\end{align}
We define the isometries
\begin{align}
    \hat{V}_A&:=A (A^\dagger A)^{-\frac{1}{2}},\quad
    \hat{V}_B:=B (B^\dagger B)^{-\frac{1}{2}},\\
    A(\alpha,\beta,\theta) &:= \sum_{\mu,\lambda\in\{0,\cdots,d-1\}}\ket{a_{\mu\lambda}}\bra{[\mu,\lambda]}\;,\\
    B(\alpha,\beta,\theta) &:= \sum_{\mu,\lambda\in\{0,\cdots,d-1\}}\ket{b_{\mu\lambda}}\bra{[\mu,\lambda]}\;,
\end{align}
where the negative matrix power is computed using the pseudo-inverse.
Following the same approach as in Eq.~\eqref{eq:Icoh_rho_Lambda}, we obtain
\begin{equation}\label{eq:Icoh_rho_Lambda_comb}
    I_c(\hat{\rho}_L, \Lambda_{\comb}) = -\tr(\hat{\omega}_A\log_2 \hat{\omega}_A)+\tr(\hat{\omega}_B\log_2 \hat{\omega}_B)\;,
\end{equation}
where $\hat{\omega}_A$ and $\hat{\omega}_B$ are represented by $dm$-by-$dm$ matrices defined by
\begin{align}
    \hat{\omega}_A(\hat{\rho}_L,\theta,\alpha,\beta) &:=   \frac{1}{\mathcal{N}_A}\sum_{\mu,\nu=0}^{d-1} (\hat{\rho}_L)_{\mu\nu} \hat{V}_A^\dagger J_{\mu\nu} \hat{V}_A \;,\\
    \hat{\omega}_B(\hat{\rho}_L,\theta,\alpha,\beta) &:=  \frac{1}{\mathcal{N}_B}\sum_{\mu,\nu=0}^{d-1} (\hat{\rho}_L)_{\mu\nu} \hat{V}_B^\dagger J^c_{\mu\nu} \hat{V}_B\;,
\end{align}
with $J_{\mu\nu}$ and $J^c_{\mu\nu}$ given by Eqs.~\eqref{eq:J_munu_comb} and \eqref{eq:J_munu_comb_c}.

\subsection{The symmetry of the coherent information}\label{app:I_coh_symm}
In this section, we offer a brief argument regarding the symmetry presented in Fig.~\ref{fig:heatmaps}, where $I_c(\mathbbm{1}/d,\Lambda|_\theta)= -I_c(\mathbbm{1}/d,\Lambda|_{\frac{\pi}{2}-\theta})$. In fact, the quantum states $\ket{\psi_\mu}$ supported on the 2D $x_1$-$x_2$ space satisfy the following transformation:
\begin{equation}
    \opU_\theta\ket{\psi_\mu} = \text{SWAP}_{1\leftrightarrow 2} \opU_{\frac{\pi}{2} - \theta} e^{i\pi \opa_1^\dagger \opa_1}\ket{\psi_{\mu}} = \text{SWAP}_{1\leftrightarrow 2} \opU_{\frac{\pi}{2} - \theta} \opD_0\ket{\psi_{d-1-\mu}}\;,
\end{equation}
where $\text{SWAP}_{1\leftrightarrow 2}$ is a unitary operation swapping modes $1$ and $2$, and $\opD_0 = \opD((d,0)/\sqrt{2})$ is independent of $\mu$ and $d$. 
Since $\opU_{\theta}\opD_0 \opU_{\theta}^\dagger = \opD_1\otimes \opD_2$ for some single-mode displacement operators $\opD_1$ and $\opD_2$, we have
$$
\hat{\omega}_A(\hat{\rho}_L,\theta,\alpha,\beta)
=
\hat{\omega}_B(\opU_{\text{flip}}\hat{\rho}_L\opU_{\text{flip}}^\dagger,\pi/2-\theta,\alpha,\beta)\;,
$$
where the unitary operation $\opU_{\text{flip}}$ is defined as $\opU_{\text{flip}} := \sum_{\mu = 0}^{d-1} \ket{\psi_{d-1-\mu}}\bra{\mu}$. Hence, we arrive at
\begin{equation}
    I_c(\hat{\rho}_L,\Lambda|_\theta)= -I_c(\opU_{\text{flip}}\hat{\rho}_L\opU_{\text{flip}}^\dagger,\Lambda|_{\frac{\pi}{2}-\theta})\;,
\end{equation}
which exhibits the symmetry when $\hat{\rho}_L$ is a maximally mixed state.

\section{Proof of equivalence between \texorpdfstring{Eq.~\eqref{eq:GGUGG_U} and Eq.~\eqref{eq:GGUGG_U_conditions}}{Eq. (GGUGG-U) and Eq. (GGUGG-U conditions)}}\label{app:proof_GGUGG_U}

Consider single-mode Gaussian unitaries $\opG_i$ and $\opG'_i$ satisfying: 
\begin{align}
    \opG_i^\dagger \opa_i \opG_i &= u_i \opa_i + v_i\opa_i^\dagger + \gamma_i\;,\\
    (\opG_i')^{\dagger} \opa_i \opG_i' &= u_i' \opa_i + v_i'\opa_i^\dagger + \gamma_i'
\end{align}
for some $u_i,v_i,\gamma_i,u_i',v_i',\gamma_i'\in\Cbb$. 
$U_\theta = \exp\big[\theta(\opa_1^\dagger \opa_2 - \opa_1\opa_2^\dagger)\big]$ is a beam-splitter unitary acting on mode $i\in\{1,2\}$.
When $\theta \in (0,\pi/2)$, we prove the following two equations
\begin{enumerate}[(1)]
    \item \begin{equation}\label{eq:GUGU}
        (\opG_1' \opG_2')^\dagger \opU_\theta (\opG_1 \opG_2) = \opU_\theta
    \end{equation}
    \item \begin{equation}\label{eq:uvgamma}
        \left\{
    \begin{aligned}
        u_1 &= u_2 = u_1' = u_2',\quad\\
        v_1 &= v_2 = v_1' = v_2',\quad\\
        \begin{pmatrix}
            \gamma_1'\\ \gamma_2'
        \end{pmatrix}&=
        \begin{pmatrix}
            \cos\theta & \sin\theta \\ 
            -\sin\theta & \cos\theta
        \end{pmatrix}
        \begin{pmatrix}
            \gamma_1\\ \gamma_2
        \end{pmatrix}\;. 
        \end{aligned}
    \right.
    \end{equation}
\end{enumerate}
are equivalent.
\begin{proof}
    First, let us prove (1)$\Rightarrow$(2).
    Denoting $R_\theta:=\begin{pmatrix}
            \cos\theta & \sin\theta \\ 
            -\sin\theta & \cos\theta
        \end{pmatrix}$, we have
    \begin{align}
        \opU_\theta^\dagger
        \begin{pmatrix}
            \opa_1 \\\opa_2  
        \end{pmatrix}
        \opU_\theta
        &=
        R_\theta
        \begin{pmatrix}
            \opa_1 \\\opa_2 
        \end{pmatrix}\; \label{eq:UaaU}\\
        \Longleftrightarrow
        (\opG_1  \opG_2)^\dagger \opU_\theta^\dagger
        \begin{pmatrix}
            \opa_1 \\\opa_2  
        \end{pmatrix}
        \opU_\theta (\opG_1  \opG_2)
        &=
        R_\theta
        \begin{pmatrix}
            u_1\opa_1 + v_1 \opa^\dagger_1 + \gamma_1 \\
            u_2\opa_2 + v_2 \opa^\dagger_2 + \gamma_2
        \end{pmatrix}\label{eq:GUaUG}\;.
    \end{align}
    Observe that Eq.~\eqref{eq:GUGU} can be rewritten as:
    \begin{equation}\label{eq:UGGU}
        (\opG_1' \opG_2')^\dagger \opU_\theta (\opG_1 \opG_2) = \opU_\theta 
        \Longleftrightarrow
        \opU_\theta (\opG_1 \opG_2) = (\opG_1' \opG_2') \opU_\theta\;,
    \end{equation} 
    Eq.~\eqref{eq:GUaUG} becomes
    \begin{align}
         \opU_\theta^\dagger(\opG_1' \opG_2')^\dagger
        \begin{pmatrix}
            \opa_1 \\\opa_2  
        \end{pmatrix}
        (\opG_1' \opG_2') \opU_\theta 
        &=
        R_\theta
        \begin{pmatrix}
            u_1\opa_1 + v_1 \opa^\dagger_1 + \gamma_1 \\
            u_2\opa_2 + v_2 \opa^\dagger_2 + \gamma_2
        \end{pmatrix}\\
        \Longleftrightarrow
         \opU_\theta^\dagger
        \begin{pmatrix}
            u_3\opa_1 + v_3 \opa^\dagger_1 + \gamma_3 \\
            u_4\opa_2 + v_4 \opa^\dagger_2 + \gamma_4
        \end{pmatrix}
        \opU_\theta 
        &=
        R_\theta
        \begin{pmatrix}
            u_1\opa_1 + v_1 \opa^\dagger_1 + \gamma_1 \\
            u_2\opa_2 + v_2 \opa^\dagger_2 + \gamma_2
        \end{pmatrix}\;.\label{eq:aaaa}
    \end{align}
    When $\theta \in (0,\pi/2)$, Eq.~\eqref{eq:UaaU} and Eq.~\eqref{eq:aaaa} imply:
    \begin{align}
        \begin{pmatrix}
        \gamma_3 \\
        \gamma_4
        \end{pmatrix}
        &=
        R_\theta
        \begin{pmatrix}
            \gamma_1 \\
            \gamma_2
        \end{pmatrix}\;,\\
        \begin{pmatrix}
            u_3&0\\ 0&u_4
        \end{pmatrix}
        R_\theta
        &=
        R_\theta
        \begin{pmatrix}
            u_1&0\\ 0&u_2
        \end{pmatrix}\;\Longrightarrow u_1 = u_2 = u_3 = u_4\;,\\
        \begin{pmatrix}
            v_3&0\\ 0&v_4
        \end{pmatrix}
        R_\theta
        &=
        R_\theta
        \begin{pmatrix}
            v_1&0\\ 0&v_2
        \end{pmatrix}\;\Longrightarrow v_1 = v_2 = v_3 = v_4\;.
    \end{align}
    
    For (2)$\Rightarrow$(1), if Eq.~\eqref{eq:uvgamma} holds, then Eq.~\eqref{eq:aaaa} is satisfied. 
    Thus, we arrive at:
    \begin{align}
        &\opU_\theta^\dagger(\opG_1' \opG_2')^\dagger
        \begin{pmatrix}
            \opa_1 \\\opa_2  
        \end{pmatrix}
        (\opG_1' \opG_2') \opU_\theta 
        =
        R_\theta
        \begin{pmatrix}
            u_1\opa_1 + v_1 \opa^\dagger_1 + \gamma_1 \\
            u_2\opa_2 + v_2 \opa^\dagger_2 + \gamma_2
        \end{pmatrix}\\
        &=
        (\opG_1 \opG_2)^\dagger
        R_\theta
        \begin{pmatrix}
            \opa_1 \\\opa_2  
        \end{pmatrix}
        (\opG_1 \opG_2)
        =
        (\opG_1 \opG_2)^\dagger
        \opU_\theta^\dagger
        \begin{pmatrix}
            \opa_1 \\\opa_2  
        \end{pmatrix}
        \opU_\theta
        (\opG_1 \opG_2)
    \end{align}
    \begin{align}
        \Longrightarrow \quad& \opU_\theta(\opG_1  \opG_2)\opU_\theta^\dagger(\opG_1' \opG_2')^\dagger
        \begin{pmatrix}
            \opa_1 \\\opa_2  
        \end{pmatrix}
        (\opG_1' \opG_2') \opU_\theta (\opG_1  \opG_2) \opU_\theta^\dagger = \begin{pmatrix}
            \opa_1 \\\opa_2  
        \end{pmatrix}\\
        \Longrightarrow\quad & \opU_\theta(\opG_1  \opG_2)\opU_\theta^\dagger(\opG_1' \opG_2')^\dagger = \mathbbm{1}\\
        \Longrightarrow\quad & (\opG_1' \opG_2')^\dagger \opU_\theta(\opG_1  \opG_2) = \opU_\theta\;.
    \end{align}
    This completes the proof of Eq.~\eqref{eq:uvgamma}.
\end{proof}

\end{document}